\documentclass[11pt]{article}
\usepackage[margin=1in]{geometry}
\usepackage{amsmath, amssymb, amsthm}
\usepackage{mathtools}
\usepackage{natbib}
\usepackage{hyperref}
\usepackage{setspace}
\usepackage{microtype}
\usepackage{booktabs}
\usepackage{bm}
\usepackage{bbm}
\usepackage{enumitem}

\usepackage{tikz}
\usepackage{bm}
\usepackage{xcolor}
\usepackage{float}

\usetikzlibrary{
    arrows.meta,
    positioning,
    calc,
    fit,
    backgrounds
}

\usepackage{caption}
\usepackage{array}
\usepackage{subcaption}
\usepackage{graphicx}
\usepackage{float}
\usepackage{tabularx}
\usepackage{multirow}
\usepackage{algorithm}
\usepackage{algorithmicx}
\usepackage{algpseudocode}  
\usepackage{siunitx}

\hypersetup{colorlinks=true,citecolor=blue,linkcolor=blue}
 
\newtheorem{theorem}{Theorem}[section]
\newtheorem{proposition}[theorem]{Proposition}
\newtheorem{lemma}[theorem]{Lemma}

\theoremstyle{definition}

\newcommand{\indep}{\perp\!\!\!\perp}

\hypersetup{
    colorlinks=true,
    citecolor=blue,   
    linkcolor=blue,
    urlcolor=blue,
}

\title{\textbf{Discrete-Time Survival Modelling with Dynamic Industrial Graphs for Mortgage Credit Risk}}

\author{
Shiqi Fang\thanks{Corresponding author. Email: \texttt{S.Fang-6@sms.ed.ac.uk}} \\
University of Edinburgh Business School \\
29 Buccleuch Place, Edinburgh EH8 9JS, UK
}
\date{}
 
\begin{document}
\maketitle

\begin{abstract}
\noindent
Mortgage credit risk may exhibit dependence across economically connected areas, even when they are geographically distant. However, current models typically represent cross-area dependence through fixed geographical neighbourhoods, potentially overlooking economic connections that extend beyond physical proximity and evolve over time. These models may fail to capture shared risks among geographically distant but economically similar areas, and cannot describe a changing dependence of risk; it therefore risks retaining economically obsolete connections while omitting newly emerging ones. This paper uses similarities of industrial composition as a measure of economic proximity and develops a discrete-time survival model with dynamic industrial graphs to capture evolving cross-area dependence in mortgage default risk. The resulting graphs are reconstructed annually, allowing both their topology and edge weights to change as regional industrial compositions evolve. Jointly combining a proper conditional autoregressive process with a first-order autoregressive process governs the temporal cross-area dependence and the temporal persistence in latent risk simultaneously. The model is estimated using Integrated Nested Laplace Approximation and evaluated on mortgage loan-level data against the static industrial graph specifications. The results show that the proposed dynamic model achieves better in-sample fit and out-of-sample discrimination, while its estimated parameters indicate substantial graph-induced cross-area dependence and temporal persistence. These findings demonstrate the value of allowing economically defined neighbourhoods to evolve when modelling mortgage default risk.

\end{abstract}


\section{Introduction}
Lenders rely on statistical models to predict not only whether a mortgage will terminate through default or full prepayment, but also when such an event is likely to occur \citep{deng2000mortgage}. Survival analysis provides a natural framework for this purpose since mortgage performance is fundamentally a time-to-event problem \citep{tong2012mixture, luo2016spline,djeundje2019dynamic}. Survival models, however, often assume that mortgage outcomes are conditionally independent once the observed covariates have been controlled for \citep{das2007common}. This assumption may be restrictive when loans located in the same or economically connected areas are exposed to common shocks and local interactions \citep{duffie2009frailty}.

The mortgage literature has primarily represented such dependence through geographic neighbourhoods. Nearby properties are exposed to similar housing markets, public services, foreclosure institutions, and neighbourhood amenities. Foreclosures and distressed sales can also depress surrounding property values by increasing local housing supply or changing expectations about neighbourhood quality \citep{anenberg2014estimates,mian2015foreclosures}. Borrowers may learn about refinancing, modification, strategic default, or foreclosure consequences through local social networks, creating additional dependence among nearby mortgage outcomes. Evidence of highly localised foreclosure effects and social interactions is provided by \citet{towe2013contagion, guiso2013determinants, gupta2019foreclosure}. 
Motivated by these, existing models have allowed the probability of default \citep{zhu2014modeling,calabrese2019mortgage}, baseline hazards \citep{calabrese2020spatial}, regression coefficients \citep{fang2021spatial}, or latent frailty terms to be related geographically, which is called spatial modelling \citep{kundig2025spatio,medina2025joint}. However, these are based on the assumption that physical closeness is a useful proxy for similarity in the economic conditions affecting mortgage performance. However, it is less satisfactory when mortgage risk is driven by common economic exposure whose boundaries do not coincide with geographic borders. Two distant regions may depend on the same manufacturing industry, energy sector, technology cluster, or agricultural activity and may consequently experience similar employment and labour markets. Conversely, two adjacent regions may have very different industrial structures and respond differently to the mortgage. 

This limitation motivates redefining neighbourhoods. Rather than selecting neighbourhoods based exclusively on geographical distance, \citet{piribauer2023beyond} estimate interregional relationships and find that sectoral production structure helps explain economic dependence among European regions. \citet{calabrese2019birds} compare spatial-only weights with weights interacting location and firm characteristics, including industry, size, and legal form, and find that this proximity captures small and medium-sized enterprises' default dependence more effectively. \citet{calabrese2023contagion} extend this through a hierarchical model that represents dependence at both industrial and geographic levels. The results show that the apparent importance of geographic clustering changes once industry-group dependence is explicitly modelled, indicating that some risks attributed to geography may instead arise from common industrial exposure. Industry should therefore not be treated solely as an additive control variable of the default predictor. It can also define the structure through which unobserved credit risk is correlated across borrowers or regions.

However, industrial neighbourhoods are not necessarily stable over time, which is different from geographic contiguity. The geographic distance between two areas changes little over the life of a mortgage, whereas their underlying industrial structures may evolve substantially. Plant openings and closures \citep{greenstone2010identifying}, sectoral expansions and contractions \citep{charles2019housing}, technological change \citep{acemoglu2020robots}, firm relocation \citep{hu2025corporate}, trade exposure \citep{autor2013china}, and longer-run structural transformation \citep{charles2019housing} can all reshape local industrial composition. As a result, areas that initially depend on similar industries may gradually diverge as one diversifies or specialises in a growing sector, while initially dissimilar areas may become more alike after operating similar activities or losing previously dominant industries. Industrial proximity is therefore intrinsically an evolving relationship rather than a permanent attribute of two areas. This temporal instability is consistent with a broader literature. Cross-industry uncertainty networks exhibit substantial variation in both their strength and connectedness over the business cycle \citep{barunik2024dynamic}, while relationships defined through economic or social distance may themselves evolve rather than remain fixed. These findings are particularly relevant when regional proximity is constructed from industrial composition, even when geographic proximity remains unchanged.

To the best of our knowledge, this distinction is not captured by current models in which the proximity and selected neighbourhoods are held fixed while the associated latent effects or dependence parameter varies over time. Such a model permits effects to strengthen or weaken along a predetermined graph on which proximity is shown, but it assumes that those edges which link the neighbours remain valid throughout the lifecycle of the mortgage. This assumption is restrictive since changes in industrial composition affect not only the magnitude of latent effects but also how they propagate across areas. A time-varying latent effect on a fixed graph can therefore describe changing risk over an unchanged graph, but it cannot describe a changing graph of risk. Consider two areas with similar industrial compositions at the beginning. A fixed industrial graph will continue to smooth their mortgage risks towards one another even if one area subsequently shifts towards services while the other shifts manufacturing. The original connection then reflects historical similarity rather than the current one. Conversely, two initially dissimilar areas may both become increasingly dependent on the same expanding industry. Their labour markets and housing markets may consequently become more closely related, yet a fixed graph cannot introduce a new connection between them. In the first case, the model retains an economically obsolete edge in the graph; in the second, it omits an economically relevant one. Neither problem can be corrected merely by allowing the latent effect to vary more flexibly over time, since the model continues to transmit dependence through the original graph.

The consequences of graph misspecification may extend beyond the estimation of latent effects. Regression coefficients may absorb unmodelled dependence, while estimated temporal effects may conflate genuine serial dependence with outdated network relationships. Predictions may consequently borrow information from areas that no longer share similar economic exposures, while overlooking emerging dependence driven by common sectoral shocks. Evidence that evolving relationships improve mortgage default prediction further highlights the importance of accounting for graph changes \citep{zandi2025attention}.
These issues also have implications for portfolio risk. Geographic diversification does not necessarily ensure economic diversification, as distant areas may share industry-specific exposures. Changes in local industrial composition may create or reduce economic risk concentration without altering the graph. A static industrial graph may therefore overlook emerging relationships or retain relationships that have weakened, potentially distorting estimates of tail risk and portfolio diversification benefits.

This study proposes a discrete-time survival model that incorporates dynamic industrial graphs to capture evolving dependence in mortgage default risk. We construct industrial graphs based on similarities in areas' industrial compositions. Each graph represents the industrial proximity between areas, with its topology identifying neighbouring areas and its edge weights measuring the strength of their industrial similarity.
A central feature of the proposed model is that the industrial graph is allowed to evolve over time. Specifically, the graph is reconstructed annually using contemporaneous industrial compositions. As local industrial composition evolves, neighbourhood relationships may emerge or disappear, and the weights assigned to existing connections may increase or decrease. These annual graphs are incorporated into a proper conditional autoregressive approach, in which the graph governs contemporaneous dependence in latent mortgage risk. At the same time, to account for persistence in the latent risk, we further introduce first-order autoregressive dynamics that link latent effects across successive years.
The proposed model distinguishes between two sources of temporal variation: changes in latent risk within areas and changes in the graphs defining dependence across areas. We additionally consider a static specification as a benchmark, in which the graph is constructed from the baseline-year industrial compositions and held fixed over time.

The paper makes five contributions. First, it extends the representation of cross-area dependence beyond geographic proximity by introducing industrial proximity as a basis for defining neighbourhoods. The proposed model embeds it to construct dependence in latent mortgage risk, allowing economically similar but geographically distant regions to be connected. Second, the study introduces a time-varying industrial graph within a discrete-time survival model, allowing both selected neighbourhoods and connections to evolve as industrial compositions change. By jointly combining annually updated graphs with a proper conditional autoregressive specification and first-order autoregressive dynamics, the model accommodates temporal variations in graphs alongside temporal persistence in latent risk. 
Third, we distinguish and compare the static specification of industrial dependence from a dynamic specification. The static specification provides an alternative in which the industrial graph and latent effect are held fixed.
Fourth, we apply the proposed model to a large panel of U.S. mortgages and compare different specifications for mortgage default. The results show that the proposed model provides the strongest overall performance, improving model fit, predictive adequacy, and discrimination.
Fifth, we offers a transferable framework for analysing risk in broad research settings where spatial dependence itself evolves.

The remainder of the paper is organised as follows. Section~\ref{sec:methodology} develops the dynamic industrial graphs and proposes the model. Section~\ref{sec:bayesian_estimation} describes the estimation process. Section~\ref{sec:evaluation} describes the evaluation framework, and the empirical analysis is presented in Section~\ref{sec:results}. Section~\ref{sec:conclusion} concludes.

\section{Methodology}
\label{sec:methodology}
\subsection{Discrete-Time Survival Model}
\label{sec:discrete_time_survival}

Consider \(N\) mortgage loans indexed by \(i=1,\ldots,N\). For loan \(i\), let \(T_i\) denote the discrete time
from origination to default and let \(C_i\) denote the corresponding right-censoring time. The observed follow-up duration and event indicator are
\begin{equation}
\widetilde T_i
=
\min(T_i,C_i),
\qquad
\delta_i
=
\mathbbm{1}(T_i\leq C_i).
\label{eq:observed_survival_data}
\end{equation}
Thus, \(\delta_i=1\) when default is observed and \(\delta_i=0\) when the loan is right-censored\footnote{Prepaid loans remain in the risk set through the month of prepayment and exit thereafter.}.

Time since origination is divided into monthly intervals
\(s=1,\ldots,\widetilde T_i\). Let
\begin{equation}
Y_{is}
=
\mathbbm{1}(\widetilde T_i=s,\delta_i=1)
\label{eq:event_indicator}
\end{equation}
denote the default indicator for loan \(i\) in duration interval \(s\).
For a defaulted loan, the observed sequence contains zeros followed by
one in the final interval; for a censored loan, it contains only zeros.

The discrete-time default hazard is
\begin{equation}
h_{is}
=
\Pr\!\left(
T_i=s
\,\middle|\,
T_i\geq s,\mathcal F_{i,s-1}
\right),
\label{eq:discrete_hazard}
\end{equation}
where \(\mathcal F_{i,s-1}\) is the information available immediately
before interval \(s\). Conditional on the linear predictor
\(\eta_{is}\),
\begin{equation}
Y_{is}\mid\eta_{is}
\sim
\operatorname{Bernoulli}(h_{is}),
\qquad
h_{is}
=
\operatorname{logit}^{-1}(\eta_{is}).
\label{eq:bernoulli_survival}
\end{equation}

Let \(\bm z_i\) contain borrower and loan characteristics. The predictor is
\begin{equation}
\eta_{is}^{(0)}
=
\beta_0
+
\bm b_s^\top\bm\alpha_f
+
\bm z_i^\top\bm\beta,
\label{eq:baseline_survival_predictor}
\end{equation}
where \(\bm b_s^\top\bm\alpha_f\) denotes the smooth duration effect represented using a penalised spline, \(\bm b_s
=\left(B_1(s),\ldots,B_{K_f}(s)\right)^\top\) denotes the spline basis evaluated at duration interval \(s\), and \(\bm\alpha_f =(\alpha_{f1},\ldots,\alpha_{fK_f})^\top\) denotes the associated spline coefficients, subject to the sum-to-zero constraint \(\sum_{s=1}^{S}\bm b_s^\top\bm\alpha_f=0\) to separate the baseline level \(\beta_0\) from its shape over duration.

Writing \(y_{is}\) for the observed value of \(Y_{is}\), the
contribution of loan \(i\) to the likelihood is
\begin{equation}
\mathcal L_i
=
\prod_{s=1}^{\widetilde T_i}
h_{is}^{y_{is}}
(1-h_{is})^{1-y_{is}},
\label{eq:individual_likelihood}
\end{equation}
and the joint likelihood is
\begin{equation}
\mathcal L
=
\prod_{i=1}^{N}\mathcal L_i.
\label{eq:likelihood}
\end{equation}


\subsection{Industrial Graph Representation}
\label{sec:industry_graphs}

The survival process is indexed by duration time \(s\), whereas industrial compositions are observed by calendar year. Let \(t\in\mathcal T=\{1,2,\cdots, T\}\) index years and let \(\mathcal A=\{1,\ldots,A\}\) denote the set of non-overlapping areas containing the properties underlying the loans. For loan \(i\), \(a_i\in\mathcal A\) identifies the area containing the associated property.

The proposed construction involves conceptually distinct quantities. The observed industrial composition of an area may change over time. Then, these changes may change the industrial graph, including both its topology and its edge weights. The industrial graph provides the cross-area dependence imposed on the latent risk.

\subsubsection{General Graph Construction}
\label{sec:industry_graph_construction}

For area \(a\) and year \(t\), let
\begin{equation}
\bm e_{a,t}
=
(e_{a1t},\ldots,e_{aKt})^\top
\in\Delta^{K-1},
\qquad
e_{akt}\geq0,
\qquad
\sum_{k=1}^{K}e_{akt}=1,
\label{eq:industry_profile}
\end{equation}
denote the industrial composition of area \(a\) across \(K\) mutually exclusive industries. Collect the area-level industrial composition in the annual profile matrix
\[
\bm P_t
=
(\bm e_{1,t},\ldots,\bm e_{A,t})^\top.
\]

For each year \(t\), industrial proximity across the \(A\) areas is represented by the weighted undirected graph
\[
\mathcal G_t
=
(\mathcal A,\mathcal E_t,\bm W_t),
\]
where \(\mathcal A\) denotes the fixed set of area vertices, \(\mathcal E_t\) denotes the set of edges representing industrial proximity between areas in year \(t\), and \(\bm W_t\) is the corresponding weighted adjacency matrix, defined as
\begin{equation}
\bm W_t
=
\mathcal W(\bm P_t;\bm\xi_I)
=
(w_{ab,t})_{a,b\in\mathcal A},
\qquad
\bm W_t=\bm W_t^\top,
\quad
w_{aa,t}=0,
\quad
w_{ab,t}\geq0,
\label{eq:annual_industry_weight_matrix}
\end{equation}
where \(\mathcal W(\,\cdot\,;\bm\xi_I)\) denote the graph construction procedure, and \(\bm\xi_I\) contains the similarity measure, neighbourhood rule, edge-weighting convention, and symmetrisation rule. These choices are held fixed across years and across the static and dynamic specifications. The edge set is determined by the positive entries of the weighted adjacency matrix:
\[
\mathcal E_t
=
\left\{
\{a,b\}\subseteq\mathcal A:
a\neq b,\ w_{ab,t}>0
\right\}.
\]
Thus, the zero--nonzero pattern of \(\bm W_t\) determines the topology of the annual graph, while the magnitude of each positive \(w_{ab,t}\) represents the strength of industrial proximity between the corresponding areas.

\subsubsection{Static and Dynamic Graphs}
\label{sec:static_dynamic_industry_graphs}

Let \(g\in\{\mathrm S,\mathrm D\}\) index the static and dynamic specifications. Let \(t_0=1\) denote the first year, also the baseline year, available in the sample.

Under the static specification, the graph constructed from the industrial composition in the baseline year is fixed throughout the observation period:
\begin{equation}
\mathcal G_t^{(\mathrm S)}
\equiv
\mathcal G_{t_0},
\qquad
\bm W_t^{(\mathrm S)}
\equiv
\bm W_{t_0}
=
\mathcal W(\bm P_{t_0};\bm\xi_I),
\qquad
t\in\mathcal T.
\label{eq:static_graph}
\end{equation}
The vertex set, topology, and edge weights are therefore identical every year.

Under the dynamic specification, the annual graph is reconstructed from the industrial composition observed in the corresponding year:
\begin{equation}
\mathcal G_t^{(\mathrm D)}
=
(\mathcal A,\mathcal E_t^{(\mathrm D)},\bm W_t^{(\mathrm D)}),
\qquad
\bm W_t^{(\mathrm D)}
=
\mathcal W(\bm P_t;\bm\xi_I),
\qquad
t\in\mathcal T.
\label{eq:dynamic_graph}
\end{equation}
Changes in industrial composition $\bm P_t$ may therefore modify \(\bm W_t^{(\mathrm D)}\), and hence graph topology, and the magnitudes of its positive entries, and hence edge strengths.

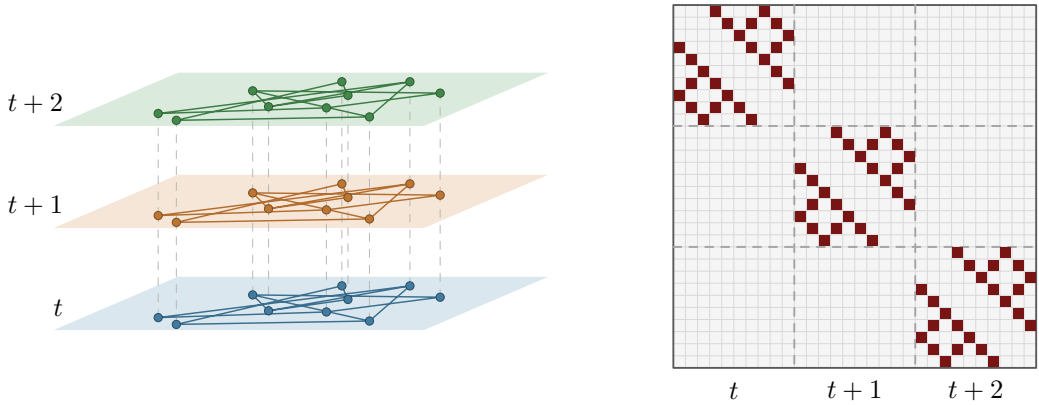
\begin{figure}[H]
\centering
\begingroup
\small
\begin{tikzpicture}[line cap=round,line join=round,font=\small]

\definecolor{layerblue}{RGB}{73,137,178}
\definecolor{layerorange}{RGB}{210,130,55}
\definecolor{layergreen}{RGB}{73,151,83}
\definecolor{matred}{RGB}{120,20,20}

\fill[layerblue,opacity=0.20]
(0,0) -- (4.90,0) -- (6.545,0.705) -- (1.645,0.705) -- cycle;

\fill[layerorange,opacity=0.20]
(0,1.35) -- (4.90,1.35) -- (6.545,2.055) -- (1.645,2.055) -- cycle;

\fill[layergreen,opacity=0.20]
(0,2.70) -- (4.90,2.70) -- (6.545,3.405) -- (1.645,3.405) -- cycle;


\coordinate (B1)  at (1.385,0.165);
\coordinate (B2)  at (2.635,0.465);
\coordinate (B3)  at (3.815,0.585);
\coordinate (B4)  at (4.715,0.585);
\coordinate (B5)  at (5.115,0.435);
\coordinate (B6)  at (1.625,0.075);
\coordinate (B7)  at (2.845,0.255);
\coordinate (B8)  at (3.610,0.240);
\coordinate (B9)  at (4.180,0.120);
\coordinate (B10) at (3.895,0.405);

\coordinate (M1)  at (1.385,1.515);
\coordinate (M2)  at (2.635,1.815);
\coordinate (M3)  at (3.815,1.935);
\coordinate (M4)  at (4.715,1.935);
\coordinate (M5)  at (5.115,1.785);
\coordinate (M6)  at (1.625,1.425);
\coordinate (M7)  at (2.845,1.605);
\coordinate (M8)  at (3.610,1.590);
\coordinate (M9)  at (4.180,1.470);
\coordinate (M10) at (3.895,1.755);

\coordinate (T1)  at (1.385,2.865);
\coordinate (T2)  at (2.635,3.165);
\coordinate (T3)  at (3.815,3.285);
\coordinate (T4)  at (4.715,3.285);
\coordinate (T5)  at (5.115,3.135);
\coordinate (T6)  at (1.625,2.775);
\coordinate (T7)  at (2.845,2.955);
\coordinate (T8)  at (3.610,2.940);
\coordinate (T9)  at (4.180,2.820);
\coordinate (T10) at (3.895,3.105);

\foreach \i in {1,...,10}{
    \draw[gray!55,dashed,line width=0.38pt]
    (B\i)--(M\i);

    \draw[gray!55,dashed,line width=0.38pt]
    (M\i)--(T\i);
}


\draw[layerblue!80!black,line width=0.55pt]
(B1)--(B4)
(B1)--(B8)
(B2)--(B5)
(B2)--(B9)
(B2)--(B7)
(B3)--(B6)
(B3)--(B10)
(B4)--(B7)
(B4)--(B9)
(B5)--(B8)
(B6)--(B9)
(B7)--(B10);

\draw[layerorange!85!black,line width=0.55pt]
(M1)--(M4)
(M1)--(M8)
(M2)--(M5)
(M2)--(M9)
(M2)--(M7)
(M3)--(M6)
(M3)--(M10)
(M4)--(M7)
(M4)--(M9)
(M5)--(M8)
(M6)--(M9)
(M7)--(M10);

\draw[layergreen!82!black,line width=0.55pt]
(T1)--(T4)
(T1)--(T8)
(T2)--(T5)
(T2)--(T9)
(T2)--(T7)
(T3)--(T6)
(T3)--(T10)
(T4)--(T7)
(T4)--(T9)
(T5)--(T8)
(T6)--(T9)
(T7)--(T10);

\foreach \i in {1,...,10}{
    \filldraw[
        fill=layerblue!90!black,
        draw=layerblue!55!black,
        line width=0.2pt]
    (B\i) circle (1.55pt);

    \filldraw[
        fill=layerorange!90!black,
        draw=layerorange!60!black,
        line width=0.2pt]
    (M\i) circle (1.55pt);

    \filldraw[
        fill=layergreen!90!black,
        draw=layergreen!60!black,
        line width=0.2pt]
    (T\i) circle (1.55pt);
}

\node[anchor=east] at (0.25,0.30) {$t$};
\node[anchor=east] at (0.25,1.65) {$t+1$};
\node[anchor=east] at (0.25,3.00) {$t+2$};


\fill[gray!8]
(8.20,4.30) rectangle (13.00,-0.50);

\foreach \k in {0,...,30}{
    \pgfmathsetmacro{\xx}{8.20+\k*0.16}
    \pgfmathsetmacro{\yy}{4.30-\k*0.16}

    \draw[gray!30,line width=0.18pt]
    (\xx,4.30)--(\xx,-0.50);

    \draw[gray!30,line width=0.18pt]
    (8.20,\yy)--(13.00,\yy);
}

\draw[black!65,line width=0.7pt]
(8.20,4.30) rectangle (13.00,-0.50);

\draw[gray!70,dashed,line width=0.75pt]
(9.80,4.30)--(9.80,-0.50);

\draw[gray!70,dashed,line width=0.75pt]
(11.40,4.30)--(11.40,-0.50);

\draw[gray!70,dashed,line width=0.75pt]
(8.20,2.70)--(13.00,2.70);

\draw[gray!70,dashed,line width=0.75pt]
(8.20,1.10)--(13.00,1.10);

\newcommand{\StaticCell}[2]{%
    \pgfmathsetmacro{\xa}{8.20+(#2-1)*0.16+0.012}
    \pgfmathsetmacro{\xb}{8.20+#2*0.16-0.012}
    \pgfmathsetmacro{\ya}{4.30-(#1-1)*0.16-0.012}
    \pgfmathsetmacro{\yb}{4.30-#1*0.16+0.012}
    \fill[matred] (\xa,\ya) rectangle (\xb,\yb);
}

\def\StaticEdges{
1/4,1/8,2/5,2/9,2/7,3/6,3/10,4/7,4/9,5/8,6/9,7/10
}

\foreach \u/\v in \StaticEdges{

    \StaticCell{\u}{\v}
    \StaticCell{\v}{\u}

    \pgfmathtruncatemacro{\uu}{\u+10}
    \pgfmathtruncatemacro{\vv}{\v+10}
    \StaticCell{\uu}{\vv}
    \StaticCell{\vv}{\uu}

    \pgfmathtruncatemacro{\uuu}{\u+20}
    \pgfmathtruncatemacro{\vvv}{\v+20}
    \StaticCell{\uuu}{\vvv}
    \StaticCell{\vvv}{\uuu}
}

\node[font=\small,align=center] at (10.60,4.75)
{};

\node at (9.00,-0.80) {$t$};
\node at (10.60,-0.80) {$t+1$};
\node at (12.20,-0.80) {$t+2$};

\end{tikzpicture}

\endgroup

\caption{
Static industrial graph across annual layers (left) and its weighted adjacency matrix (right). Each layer corresponds to one year and contains the same area vertex. Coloured within-layer edges represent connections, while vertical dashed lines identify the same areas across years.
}
\label{fig:static_industry_structure}
\end{figure}

\begin{figure}[H]
\centering
\begingroup
\small
\begin{tikzpicture}[line cap=round,line join=round,font=\small]

\definecolor{layerblue}{RGB}{73,137,178}
\definecolor{layerorange}{RGB}{210,130,55}
\definecolor{layergreen}{RGB}{73,151,83}
\definecolor{matred}{RGB}{120,20,20}

\fill[layerblue,opacity=0.20]
(0,0) -- (4.90,0) -- (6.545,0.705) -- (1.645,0.705) -- cycle;

\fill[layerorange,opacity=0.20]
(0,1.35) -- (4.90,1.35) -- (6.545,2.055) -- (1.645,2.055) -- cycle;

\fill[layergreen,opacity=0.20]
(0,2.70) -- (4.90,2.70) -- (6.545,3.405) -- (1.645,3.405) -- cycle;


\coordinate (B1)  at (1.385,0.165);
\coordinate (B2)  at (2.635,0.465);
\coordinate (B3)  at (3.815,0.585);
\coordinate (B4)  at (4.715,0.585);
\coordinate (B5)  at (5.115,0.435);
\coordinate (B6)  at (1.625,0.075);
\coordinate (B7)  at (2.845,0.255);
\coordinate (B8)  at (3.610,0.240);
\coordinate (B9)  at (4.180,0.120);
\coordinate (B10) at (3.895,0.405);

\coordinate (M1)  at (1.385,1.515);
\coordinate (M2)  at (2.635,1.815);
\coordinate (M3)  at (3.815,1.935);
\coordinate (M4)  at (4.715,1.935);
\coordinate (M5)  at (5.115,1.785);
\coordinate (M6)  at (1.625,1.425);
\coordinate (M7)  at (2.845,1.605);
\coordinate (M8)  at (3.610,1.590);
\coordinate (M9)  at (4.180,1.470);
\coordinate (M10) at (3.895,1.755);

\coordinate (T1)  at (1.385,2.865);
\coordinate (T2)  at (2.635,3.165);
\coordinate (T3)  at (3.815,3.285);
\coordinate (T4)  at (4.715,3.285);
\coordinate (T5)  at (5.115,3.135);
\coordinate (T6)  at (1.625,2.775);
\coordinate (T7)  at (2.845,2.955);
\coordinate (T8)  at (3.610,2.940);
\coordinate (T9)  at (4.180,2.820);
\coordinate (T10) at (3.895,3.105);

\foreach \i in {1,...,10}{
    \draw[gray!55,dashed,line width=0.38pt]
    (B\i)--(M\i);

    \draw[gray!55,dashed,line width=0.38pt]
    (M\i)--(T\i);
}


\draw[layerblue!80!black,line width=0.55pt] (B1)--(B4);
\draw[layerblue!80!black,line width=0.55pt] (B1)--(B8);
\draw[layerblue!80!black,line width=0.55pt] (B2)--(B5);
\draw[layerblue!80!black,line width=0.55pt] (B2)--(B9);
\draw[layerblue!80!black,line width=0.55pt] (B2)--(B7);
\draw[layerblue!80!black,line width=0.55pt] (B3)--(B6);
\draw[layerblue!80!black,line width=0.55pt] (B3)--(B10);
\draw[layerblue!80!black,line width=0.55pt] (B4)--(B7);
\draw[layerblue!80!black,line width=0.55pt] (B4)--(B9);
\draw[layerblue!80!black,line width=0.55pt] (B5)--(B8);
\draw[layerblue!80!black,line width=0.55pt] (B6)--(B9);
\draw[layerblue!80!black,line width=0.55pt] (B7)--(B10);

\draw[layerorange!85!black,line width=0.55pt] (M1)--(M5);
\draw[layerorange!85!black,line width=0.55pt] (M1)--(M9);
\draw[layerorange!85!black,line width=0.55pt] (M2)--(M6);
\draw[layerorange!85!black,line width=0.55pt] (M2)--(M10);
\draw[layerorange!85!black,line width=0.55pt] (M3)--(M7);
\draw[layerorange!85!black,line width=0.55pt] (M3)--(M9);
\draw[layerorange!85!black,line width=0.55pt] (M4)--(M8);
\draw[layerorange!85!black,line width=0.55pt] (M4)--(M10);
\draw[layerorange!85!black,line width=0.55pt] (M5)--(M7);
\draw[layerorange!85!black,line width=0.55pt] (M6)--(M8);
\draw[layerorange!85!black,line width=0.55pt] (M7)--(M9);
\draw[layerorange!85!black,line width=0.55pt] (M8)--(M10);

\draw[layergreen!82!black,line width=0.55pt] (T1)--(T6);
\draw[layergreen!82!black,line width=0.55pt] (T1)--(T10);
\draw[layergreen!82!black,line width=0.55pt] (T2)--(T4);
\draw[layergreen!82!black,line width=0.55pt] (T2)--(T8);
\draw[layergreen!82!black,line width=0.55pt] (T3)--(T5);
\draw[layergreen!82!black,line width=0.55pt] (T3)--(T9);
\draw[layergreen!82!black,line width=0.55pt] (T4)--(T7);
\draw[layergreen!82!black,line width=0.55pt] (T5)--(T10);
\draw[layergreen!82!black,line width=0.55pt] (T6)--(T9);
\draw[layergreen!82!black,line width=0.55pt] (T7)--(T8);
\draw[layergreen!82!black,line width=0.55pt] (T7)--(T10);
\draw[layergreen!82!black,line width=0.55pt] (T2)--(T9);

\foreach \i in {1,...,10}{
    \filldraw[
        fill=layerblue!90!black,
        draw=layerblue!55!black,
        line width=0.2pt]
    (B\i) circle (1.55pt);

    \filldraw[
        fill=layerorange!90!black,
        draw=layerorange!60!black,
        line width=0.2pt]
    (M\i) circle (1.55pt);

    \filldraw[
        fill=layergreen!90!black,
        draw=layergreen!60!black,
        line width=0.2pt]
    (T\i) circle (1.55pt);
}

\node[anchor=east] at (0.25,0.30) {$t$};
\node[anchor=east] at (0.25,1.65) {$t+1$};
\node[anchor=east] at (0.25,3.00) {$t+2$};


\fill[gray!8]
(8.20,4.30) rectangle (13.00,-0.50);

\foreach \k in {0,...,30}{
    \pgfmathsetmacro{\xx}{8.20+\k*0.16}
    \pgfmathsetmacro{\yy}{4.30-\k*0.16}

    \draw[gray!30,line width=0.18pt]
    (\xx,4.30)--(\xx,-0.50);

    \draw[gray!30,line width=0.18pt]
    (8.20,\yy)--(13.00,\yy);
}

\draw[black!65,line width=0.7pt]
(8.20,4.30) rectangle (13.00,-0.50);

\draw[gray!70,dashed,line width=0.75pt]
(9.80,4.30)--(9.80,-0.50);

\draw[gray!70,dashed,line width=0.75pt]
(11.40,4.30)--(11.40,-0.50);

\draw[gray!70,dashed,line width=0.75pt]
(8.20,2.70)--(13.00,2.70);

\draw[gray!70,dashed,line width=0.75pt]
(8.20,1.10)--(13.00,1.10);

\newcommand{\DynamicCell}[2]{%
    \pgfmathsetmacro{\xa}{8.20+(#2-1)*0.16+0.012}
    \pgfmathsetmacro{\xb}{8.20+#2*0.16-0.012}
    \pgfmathsetmacro{\ya}{4.30-(#1-1)*0.16-0.012}
    \pgfmathsetmacro{\yb}{4.30-#1*0.16+0.012}
    \fill[matred] (\xa,\ya) rectangle (\xb,\yb);
}

\def\EdgesOne{
1/4,1/8,2/5,2/9,2/7,3/6,3/10,4/7,4/9,5/8,6/9,7/10
}

\def\EdgesTwo{
1/5,1/9,2/6,2/10,3/7,3/9,4/8,4/10,5/7,6/8,7/9,8/10
}

\def\EdgesThree{
1/6,1/10,2/4,2/8,3/5,3/9,4/7,5/10,6/9,7/8,7/10,2/9
}

\foreach \u/\v in \EdgesOne{
    \DynamicCell{\u}{\v}
    \DynamicCell{\v}{\u}
}

\foreach \u/\v in \EdgesTwo{
    \pgfmathtruncatemacro{\uu}{\u+10}
    \pgfmathtruncatemacro{\vv}{\v+10}

    \DynamicCell{\uu}{\vv}
    \DynamicCell{\vv}{\uu}
}

\foreach \u/\v in \EdgesThree{
    \pgfmathtruncatemacro{\uuu}{\u+20}
    \pgfmathtruncatemacro{\vvv}{\v+20}

    \DynamicCell{\uuu}{\vvv}
    \DynamicCell{\vvv}{\uuu}
}

\node[font=\small,align=center] at (10.60,4.75)
{};

\node at (9.00,-0.80) {$t$};
\node at (10.60,-0.80) {$t+1$};
\node at (12.20,-0.80) {$t+2$};

\end{tikzpicture}
\endgroup

\caption{
Dynamic industrial graph across annual layers (left) and its weighted adjacency matrix (right). Each layer corresponds to one year and contains the same area vertex. Coloured within-layer edges represent connections, while vertical dashed lines identify the same areas across years.
}
\label{fig:dynamic_industry_structure}
\end{figure}
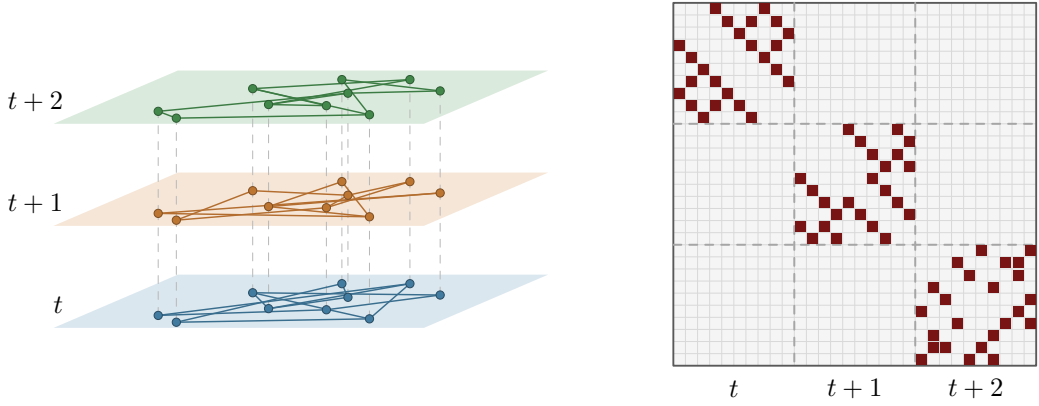

By construction,
\[
\mathcal G_{t_0}^{(\mathrm S)}
=
\mathcal G_{t_0}^{(\mathrm D)}.
\]
The two specifications therefore begin from the same industrial graph. The static specification repeatedly uses the baseline graph, whereas the dynamic updates the graphs from contemporaneous industrial composition. Fig.~\ref{fig:static_industry_structure} and~\ref{fig:dynamic_industry_structure} illustrate this distinction.

\subsection{Graph-Structured Latent Dependence}
\label{sec:graph_induced_dependence}
The graph is not merely a descriptive representation. Its topology and edge weights determine the contemporaneous cross-area dependence that is embedded in the model.

\subsubsection{Cross-Area Dependence}
\label{sec:graph_laplacian}

For specification \(g\in\{\mathrm S,\mathrm D\}\), define the degree
matrix
\[
\bm D_t^{(g)}
=
\operatorname{diag}
\left(
\bm W_t^{(g)}\bm 1_A
\right).
\]
The \(a\)th diagonal element is the total industrial-proximity weight linking area \(a\) to the remaining areas in year \(t\).

The corresponding weighted graph Laplacian is
\begin{equation}
\bm L_t^{(g)}
=
\bm D_t^{(g)}
-
\bm W_t^{(g)}.
\label{eq:annual_graph_laplacian}
\end{equation}

For any area-level vector \(\bm x=(x_1,\ldots,x_A)^\top\in\mathbb R^A\), the associated quadratic form satisfies
\begin{equation}
\bm x^\top
\bm L_t^{(g)}
\bm x
=
\sum_{a<b}
w_{ab,t}^{(g)}
(x_a-x_b)^2.
\label{eq:graph_laplacian_quadratic_form}
\end{equation}
Hence, the graph topology determines which areas are treated as neighbours and connected, while the edge weights determine the strength of that connection. Differences between areas connected by larger industrial-proximity weights receive larger quadratic penalties. The proof is provided in Appendix~\ref{lem:laplacian_psd}.

\subsubsection{Scaling and Proper Graph Structure}
\label{sec:graph_scaling}

The scale implied by a graph Laplacian depends on both its topology and its edge weights. Therefore, graphs with different connectivity patterns may imply different levels of prior variability even when multiplied by the same precision scale parameter. This is particularly important under the dynamic specification, where the annual graph itself may change.

To place the graphs on a comparable scale, we therefore first standardise the graph. Following \citet{sorbye2014scaling}, for a connected graph define
\begin{equation}
c_t^{(g)}
=
\exp
\left\{
\frac{1}{A}
\sum_{a=1}^{A}
\log
\left[
\left\{
\left(
\bm L_t^{(g)}
\right)^{-}
\right\}_{aa}
\right]
\right\},
\qquad
\widetilde{\bm L}_t^{(g)}
=
c_t^{(g)}
\bm L_t^{(g)},
\label{eq:scaled_graph_structure}
\end{equation}
where
\((\bm L_t^{(g)})^{-}\) denotes its Moore-Penrose generalized inverse \citep{keefe2018formal}. The scaled matrix
\(\widetilde{\bm L}_t^{(g)}\) preserves the topology and relative edge weight of the original graph while placing the graph-induced variability on a common generalised-variance scale. The proof is presented in Appendix~\ref{prop:graph_scaling}.

A graph Laplacian is singular, however, and the graph should shape area-level variation without requiring that all such
variation be explained through the graph-induced cross-area dependence. We therefore combine the scaled graph with an unstructured component:
\begin{equation}
\widehat{\bm L}_t^{(g)}(\lambda)
=
(1-\lambda)I_A
+
\lambda
\widetilde{\bm L}_t^{(g)},
\qquad
0\leq\lambda<1.
\label{eq:proper_graph_structure}
\end{equation}
The structure weight \(\lambda\) controls the balance between unstructured and graph-induced cross-area dependence. Since
\(\widetilde{\bm L}_t^{(g)}\) is symmetric positive semidefinite, \(\widehat{\bm L}_t^{(g)}(\lambda)\) is symmetric positive definite for \(\lambda<1\), see Appendix~\ref{prop:innovation_properness}.

Although \(\widetilde{\bm L}_t^{(g)}\) has already been standardised and \(\widehat{\bm L}_t^{(g)}(\lambda)\) is proper, its implied marginal variance scale may still depend on the particular graph and on \(\lambda\). Define
\begin{equation}
r_t^{(g)}(\lambda)
=
\exp
\left\{
\frac{1}{A}
\sum_{a=1}^{A}
\log
\left[
\left\{
\widehat{\bm L}_t^{(g)}(\lambda)^{-1}
\right\}_{aa}
\right]
\right\},
\end{equation}
and set
\begin{equation}
\bm R_t^{(g)}(\lambda)
=
r_t^{(g)}(\lambda)
\widehat{\bm L}_t^{(g)}(\lambda).
\label{eq:annual_graph_precision}
\end{equation}

The two scaling steps serve different purposes. The factor \(c_t^{(g)}\) standardises the graph before it is mixed with the
unstructured component, giving \(\lambda\) a comparable interpretation across graph structures. The factor \(r_t^{(g)}(\lambda)\) then standardises the resulting proper precision structure, placing its implied marginal variability on a common scale. \(\bm R_t^{(g)}(\lambda)\) is therefore the final proper, scale-standardised cross-area precision matrix through which the graph enters the model; see Appendix~\ref{prop:innovation_properness}-\ref{prop:tau_interpretation}.


\subsection{Static and Dynamic Latent Effects}
\subsubsection{Static Latent Effect}
\label{sec:static_latent_effect}

Under the static specification, the industrial graph is fixed, so that
\[
\bm R_t^{(\mathrm S)}(\lambda)
\equiv
\bm R_{t_0}^{(\mathrm S)}(\lambda)
\]
The graph-induced cross-area dependence is therefore represented by a single latent effect
\[
\bm u
=
(u_1,\ldots,u_A)^\top.
\]
\(u_a \in \bm u\) represents persistent residual variation in mortgage risk associated with area \(a\) after accounting for the observed covariates and the duration effect. Conditional on the precision scale $\tau_u$ and parameter $\lambda$, we assume
\begin{equation}
\bm u
\mid
\tau_u,\lambda
\sim
\mathcal N_A
\left[
\bm 0,
\left\{
\bm Q_u^{(\mathrm S)}
\right\}^{-1}
\right],
\label{eq:static_latent_effect}
\end{equation}
where $\bm Q_u^{(\mathrm S)}=\tau_u \bm R_{t_0}^{(\mathrm S)}(\lambda)$ and $\tau_u>0$ controls the scale of the latent effect.

\subsubsection{Dynamic Latent Effect}
\label{sec:graph_temporal_process}

The dynamic specification, in contrast, incorporates an additional form of dependence, distinct from the cross-area dependence. Across years, persistence in the latent effect is captured through an autoregressive formulation.

For each year \(t\in\mathcal T\), let
\[
\bm\phi_t
=
(\phi_{1,t},\ldots,\phi_{A,t})^\top
\]
denote the latent effect of mortgage risk across areas. \(\phi_{a,t} \in \bm\phi_t\) captures residual area-year variation in mortgage risk associated with industrial conditions after accounting for the observed covariates and the duration effect.

Regional economic conditions are persistent rather than independently redrawn from one year to the next. Local employment, labour-market conditions, and their associated exposure to sectoral shocks typically evolve gradually, so an area experiencing elevated industry-related risk in one year is likely to retain part of that risk in the following year. Treating \(\bm\phi_1,\ldots,\bm\phi_T\) as independent would discard this temporal continuity and would attribute all year-to-year changes to new contemporaneous shocks. We therefore model the latent field as a first-order autoregressive process,
\begin{equation}
\bm\phi_t
=
\rho\bm\phi_{t-1}
+
\bm\varepsilon_t,
\qquad
t=2,\ldots,T,
\qquad
|\rho|<1.
\label{eq:industry_ar1_state}
\end{equation}
The parameter \(\rho\) measures temporal persistence. The innovation \(\bm\varepsilon_t\) represents new latent area-level risk variation in year \(t\) after removing the part inherited from year \(t-1\). The role of the contemporaneous industrial graph is to determine how these new innovations are related across areas.

Conditional on the model parameters, we assume that the initial latent effect $\bm\phi_1$ and the subsequent innovations
$\bm\varepsilon_2,\ldots,\bm\varepsilon_T$
are mutually independent
\begin{equation}
\bm\varepsilon_t
\mid
\tau_\phi,\rho,\lambda
\sim
\mathcal N_A
\left[
\bm0,
\left\{
\tau_\phi
\bm R_t^{(\mathrm D)}(\lambda)
\right\}^{-1}
\right],
\qquad
t=2,\ldots,T.
\label{eq:annual_innovation_distribution}
\end{equation}

\(\tau_\phi>0\) controls the overall magnitude of those innovations. Thus, areas that are more strongly connected in the contemporaneous graph are encouraged to experience more similar new latent risk variation, while the unstructured component in $\bm R_t^{(\mathrm D)}(\lambda)$ permits area-specific departures from this pattern.

The autoregressive specification in Eq.\eqref{eq:industry_ar1_state} determines how the latent effects evolves yearly, but it does not by itself specify the distribution of the effect in the initial year. We then require:
\begin{equation}
\bm\phi_1
\mid
\tau_\phi,\rho,\lambda
\sim
\mathcal N_A
\left[
\bm0,
\left\{
\tau_\phi(1-\rho^2)
\bm R_1^{(\mathrm D)}(\lambda)
\right\}^{-1}
\right].
\label{eq:industry_initial_distribution}
\end{equation}

Equivalently, define the transformed innovations as
\begin{equation}
\bm\varepsilon_1
=
\sqrt{1-\rho^2}\,\bm\phi_1,
\qquad
\bm\varepsilon_t
=
\bm\phi_t
-
\rho\bm\phi_{t-1},
\quad
t=2,\ldots,T.
\label{eq:transformed_industry_innovations}
\end{equation}
Then,
\begin{equation}
\bm\varepsilon_t
\mid
\tau_\phi,\lambda
\sim
\mathcal N_A
\left[
\bm0,
\left\{
\tau_\phi
\bm R_t^{(\mathrm D)}(\lambda)
\right\}^{-1}
\right],
\qquad
t=1,\ldots,T.
\label{eq:common_innovation_distribution}
\end{equation}
Hence, the initial state and all subsequent innovations are expressed in the same formulation. The equivalence is established in Appendix~\ref{prop:unified_innovation}.

\paragraph{Joint Dependence Representation}
\label{sec:joint_industry_representation}

To characterise the joint dependence arising from the interplay between cross-area dependence and the temporal evolution of latent effects, we stack the annual latent effects and transformed innovations as
\[
\bm\phi
=
(\bm\phi_1^\top,\ldots,\bm\phi_T^\top)^\top,
\qquad
\bm\varepsilon
=
(\bm\varepsilon_1^\top,\ldots,\bm\varepsilon_T^\top)^\top
\in\mathbb R^{AT}.
\]

Conditional on the parameters, the transformed innovations are independent across years, while their contemporaneous cross-area dependence is determined by the annual graph. The annual precision matrix can therefore be collected as
\begin{equation}
\mathcal R^{(\mathrm D)}(\lambda)
=
\operatorname{blockdiag}
\left\{
\bm R_1^{(\mathrm D)}(\lambda),
\ldots,
\bm R_T^{(\mathrm D)}(\lambda)
\right\}.
\label{eq:block_graph_structure}
\end{equation}

The matrix \(\mathcal R^{(\mathrm D)}(\lambda)\in\mathbb R^{AT\times AT}\) contains contemporaneous cross-area dependence but no direct temporal evolution. Its \(t\)th diagonal \(A\times A\) block, \(\bm R_t^{(\mathrm D)}(\lambda)\), determines how the innovations of the \(A\) areas are related within year \(t\), as in Eq.\eqref{eq:common_innovation_distribution}.

Temporal evolution enters separately through the autoregressive approach. Define
\(\mathbb B_\rho=B_\rho\otimes I_A\), where
\begin{equation}
B_\rho
=
\begin{pmatrix}
\sqrt{1-\rho^2} & 0      & \cdots & 0 \\
-\rho           & 1      & \ddots & \vdots \\
0               & \ddots & \ddots & 0 \\
0               & \cdots & -\rho  & 1
\end{pmatrix}.
\label{eq:temporal_innovation_matrix}
\end{equation}
The transformed innovations can then be written as
\begin{equation}
\bm\varepsilon
=
\mathbb B_\rho\bm\phi.
\label{eq:transformed_innovation}
\end{equation}
The proof is presented in Appendix~\ref{lem:stacked_innovation_transformation}.

The roles of \(\mathcal R^{(\mathrm D)}(\lambda)\) and \(\mathbb B_\rho\) are therefore distinct. The former determines
contemporaneous cross-area dependence in the innovations within each year, thereby affecting the latent effects, whereas the latter determines how latent effects are propagated across consecutive years.

Combining the two gives the joint precision matrix of the dynamic latent field:
\begin{equation}
\bm Q_\phi^{(\mathrm D)}
=
\tau_\phi
\mathbb B_\rho^\top
\mathcal R^{(\mathrm D)}(\lambda)
\mathbb B_\rho.
\label{eq:joint_industry_precision}
\end{equation}
Accordingly,
\begin{equation}
\bm\phi
\mid
\tau_\phi,\rho,\lambda
\sim
\mathcal N_{AT}
\left[
\bm0,
\left\{
\bm Q_\phi^{(\mathrm D)}
\right\}^{-1}
\right],
\label{eq:joint_industry_prior}
\end{equation}
before imposing the identification constraints introduced in Section~\ref{sec:identification}. The proof is presented in Appendix~\ref{prop:joint_precision}.

\paragraph{Dynamic Nonseparability}
\label{sec:static_dynamic_joint_structures}

Under the dynamic specification, since the cross-area dependence changes across years, this joint precision matrix generally cannot be reduced to the Kronecker product of a single temporal matrix and a single fixed cross-area matrix. The dynamic specification is therefore generally non-separable. The corresponding result is established in Appendix~\ref{prop:dynamic_nonseparability}.

\begin{figure}[H]
\centering
\resizebox{\textwidth}{!}{%
\begin{tikzpicture}[
font=\small,
line cap=round,
line join=round,
>=Latex,
myarrow/.style={
  -{Latex[length=3mm]},
  draw=gray!70,
  line width=0.9pt
}]


\definecolor{layerblue}{RGB}{73,137,178}
\definecolor{layerorange}{RGB}{210,130,55}
\definecolor{layergreen}{RGB}{73,151,83}


\draw[
  rounded corners=3pt,
  draw=blue!55,
  fill=blue!5,
  line width=0.9pt
]
(0.20,9.00) rectangle (9.40,14.30);

\node[
  font=\bfseries\large,
  anchor=west
]
at (0.65,13.90)
{1. Annual industrial profiles};

\draw[
  rounded corners=2pt,
  draw=blue!35,
  fill=white,
  line width=0.7pt
]
(0.55,9.35) rectangle (9.05,13.50);

\draw[
  rounded corners=3pt,
  draw=blue!68!black,
  dashed,
  dash pattern=on 3pt off 2pt,
  line width=0.75pt
]
(0.75,9.52) rectangle (8.85,13.33);

\node[
  fill=white,
  inner xsep=4pt,
  inner ysep=1pt,
  font=\scriptsize,
  text=blue!70!black
]
at (4.80,13.33)
{One profile matrix per year: $t=1,\ldots,T$};

\begin{scope}[xshift=0.65cm]

\node[font=\Large]
at (4.10,12.88)
{$
\bm P_t =
(\bm e_{1,t},\ldots,\bm e_{A,t})^\top
$};

\node at (2.75,12.35) {Sec. 1};
\node at (3.65,12.35) {Sec. 2};
\node at (4.55,12.35) {$\cdots$};
\node at (5.45,12.35) {Sec. $K$};

\node[anchor=east] at (2.20,11.85) {Area 1};
\node[anchor=east] at (2.20,11.20) {Area 2};
\node[anchor=east] at (2.20,10.55) {$\vdots$};
\node[anchor=east] at (2.20,9.90) {Area $A$};

\fill[blue!20] (2.40,11.58) rectangle (3.10,12.05);
\fill[blue!55] (3.30,11.58) rectangle (4.00,12.05);
\fill[blue!15] (4.20,11.58) rectangle (4.90,12.05);
\fill[blue!35] (5.10,11.58) rectangle (5.80,12.05);

\fill[blue!50] (2.40,10.93) rectangle (3.10,11.40);
\fill[blue!20] (3.30,10.93) rectangle (4.00,11.40);
\fill[blue!38] (4.20,10.93) rectangle (4.90,11.40);
\fill[blue!65] (5.10,10.93) rectangle (5.80,11.40);

\fill[blue!15] (2.40,10.28) rectangle (3.10,10.75);
\fill[blue!70] (3.30,10.28) rectangle (4.00,10.75);
\fill[blue!10] (4.20,10.28) rectangle (4.90,10.75);
\fill[blue!25] (5.10,10.28) rectangle (5.80,10.75);

\fill[blue!40] (2.40,9.63) rectangle (3.10,10.10);
\fill[blue!30] (3.30,9.63) rectangle (4.00,10.10);
\fill[blue!55] (4.20,9.63) rectangle (4.90,10.10);
\fill[blue!20] (5.10,9.63) rectangle (5.80,10.10);

\end{scope}


\draw[
  -{Latex[length=3mm]},
  draw=gray!65,
  line width=1.3pt
]
(9.40,11.65)--(10.20,11.65);


\draw[
  rounded corners=3pt,
  draw=green!55!black,
  fill=green!5,
  line width=0.9pt
]
(10.20,9.00) rectangle (21.40,14.30);

\draw[
  rounded corners=2pt,
  draw=green!45!black,
  fill=white,
  line width=0.7pt
]
(10.50,9.35) rectangle (21.10,13.55);

\node[
  font=\bfseries\large,
  anchor=west
]
at (10.90,13.90)
{2. Annual industrial graph};

\node[font=\small]
at (15.80,13.34)
{$
\mathcal G_t^{(\mathrm D)}
=
(\mathcal A,\mathcal E_t^{(\mathrm D)},
\bm W_t^{(\mathrm D)})
$};

\def\YearBlueEdges{1/4,1/8,2/5,2/9,3/6,3/10,4/7,5/8,6/9,7/10}


\begin{scope}[shift={(12.60,9.55)}]

\clip (-1.22,-0.14) rectangle (7.82,3.94);

\fill[layerblue,opacity=0.20]
(0,0)--(4.90,0)--(6.545,0.705)--(1.645,0.705)--cycle;

\fill[layerorange,opacity=0.20]
(0,1.35)--(4.90,1.35)--(6.545,2.055)--(1.645,2.055)--cycle;

\fill[layergreen,opacity=0.20]
(0,2.70)--(4.90,2.70)--(6.545,3.405)--(1.645,3.405)--cycle;


\coordinate (modelB1) at (1.385,0.165);
\coordinate (modelB2) at (2.635,0.465);
\coordinate (modelB3) at (3.815,0.585);
\coordinate (modelB4) at (4.715,0.585);
\coordinate (modelB5) at (5.115,0.435);
\coordinate (modelB6) at (1.625,0.075);
\coordinate (modelB7) at (2.845,0.255);
\coordinate (modelB8) at (3.610,0.240);
\coordinate (modelB9) at (4.180,0.120);
\coordinate (modelB10) at (3.895,0.405);


\coordinate (modelM1) at (1.385,1.515);
\coordinate (modelM2) at (2.635,1.815);
\coordinate (modelM3) at (3.815,1.935);
\coordinate (modelM4) at (4.715,1.935);
\coordinate (modelM5) at (5.115,1.785);
\coordinate (modelM6) at (1.625,1.425);
\coordinate (modelM7) at (2.845,1.605);
\coordinate (modelM8) at (3.610,1.590);
\coordinate (modelM9) at (4.180,1.470);
\coordinate (modelM10) at (3.895,1.755);


\coordinate (modelT1) at (1.385,2.865);
\coordinate (modelT2) at (2.635,3.165);
\coordinate (modelT3) at (3.815,3.285);
\coordinate (modelT4) at (4.715,3.285);
\coordinate (modelT5) at (5.115,3.135);
\coordinate (modelT6) at (1.625,2.775);
\coordinate (modelT7) at (2.845,2.955);
\coordinate (modelT8) at (3.610,2.940);
\coordinate (modelT9) at (4.180,2.820);
\coordinate (modelT10) at (3.895,3.105);


\draw[gray!55,dashed,line width=0.38pt] (modelB1)--(modelM1);
\draw[gray!55,dashed,line width=0.38pt] (modelM1)--(modelT1);

\draw[gray!55,dashed,line width=0.38pt] (modelB2)--(modelM2);
\draw[gray!55,dashed,line width=0.38pt] (modelM2)--(modelT2);

\draw[gray!55,dashed,line width=0.38pt] (modelB3)--(modelM3);
\draw[gray!55,dashed,line width=0.38pt] (modelM3)--(modelT3);

\draw[gray!55,dashed,line width=0.38pt] (modelB4)--(modelM4);
\draw[gray!55,dashed,line width=0.38pt] (modelM4)--(modelT4);

\draw[gray!55,dashed,line width=0.38pt] (modelB5)--(modelM5);
\draw[gray!55,dashed,line width=0.38pt] (modelM5)--(modelT5);

\draw[gray!55,dashed,line width=0.38pt] (modelB6)--(modelM6);
\draw[gray!55,dashed,line width=0.38pt] (modelM6)--(modelT6);

\draw[gray!55,dashed,line width=0.38pt] (modelB7)--(modelM7);
\draw[gray!55,dashed,line width=0.38pt] (modelM7)--(modelT7);

\draw[gray!55,dashed,line width=0.38pt] (modelB8)--(modelM8);
\draw[gray!55,dashed,line width=0.38pt] (modelM8)--(modelT8);

\draw[gray!55,dashed,line width=0.38pt] (modelB9)--(modelM9);
\draw[gray!55,dashed,line width=0.38pt] (modelM9)--(modelT9);

\draw[gray!55,dashed,line width=0.38pt] (modelB10)--(modelM10);
\draw[gray!55,dashed,line width=0.38pt] (modelM10)--(modelT10);


\draw[layerblue!80!black,line width=0.55pt] (modelB1)--(modelB4);
\draw[layerblue!80!black,line width=0.55pt] (modelB1)--(modelB8);
\draw[layerblue!80!black,line width=0.55pt] (modelB2)--(modelB5);
\draw[layerblue!80!black,line width=0.55pt] (modelB2)--(modelB9);
\draw[layerblue!80!black,line width=0.55pt] (modelB3)--(modelB6);
\draw[layerblue!80!black,line width=0.55pt] (modelB3)--(modelB10);
\draw[layerblue!80!black,line width=0.55pt] (modelB4)--(modelB7);
\draw[layerblue!80!black,line width=0.55pt] (modelB5)--(modelB8);
\draw[layerblue!80!black,line width=0.55pt] (modelB6)--(modelB9);
\draw[layerblue!80!black,line width=0.55pt] (modelB7)--(modelB10);


\draw[layerorange!80!black,line width=0.55pt] (modelM1)--(modelM5);
\draw[layerorange!80!black,line width=0.55pt] (modelM1)--(modelM9);
\draw[layerorange!80!black,line width=0.55pt] (modelM2)--(modelM6);
\draw[layerorange!80!black,line width=0.55pt] (modelM2)--(modelM10);
\draw[layerorange!80!black,line width=0.55pt] (modelM3)--(modelM7);
\draw[layerorange!80!black,line width=0.55pt] (modelM3)--(modelM9);
\draw[layerorange!80!black,line width=0.55pt] (modelM4)--(modelM8);
\draw[layerorange!80!black,line width=0.55pt] (modelM4)--(modelM10);
\draw[layerorange!80!black,line width=0.55pt] (modelM5)--(modelM7);
\draw[layerorange!80!black,line width=0.55pt] (modelM6)--(modelM8);
\draw[layerorange!80!black,line width=0.55pt] (modelM7)--(modelM9);
\draw[layerorange!80!black,line width=0.55pt] (modelM8)--(modelM10);


\draw[layergreen!80!black,line width=0.55pt] (modelT1)--(modelT6);
\draw[layergreen!80!black,line width=0.55pt] (modelT1)--(modelT10);
\draw[layergreen!80!black,line width=0.55pt] (modelT2)--(modelT4);
\draw[layergreen!80!black,line width=0.55pt] (modelT2)--(modelT8);
\draw[layergreen!80!black,line width=0.55pt] (modelT3)--(modelT5);
\draw[layergreen!80!black,line width=0.55pt] (modelT3)--(modelT9);
\draw[layergreen!80!black,line width=0.55pt] (modelT4)--(modelT7);
\draw[layergreen!80!black,line width=0.55pt] (modelT5)--(modelT10);
\draw[layergreen!80!black,line width=0.55pt] (modelT6)--(modelT9);
\draw[layergreen!80!black,line width=0.55pt] (modelT7)--(modelT8);
\draw[layergreen!80!black,line width=0.55pt] (modelT7)--(modelT10);
\draw[layergreen!80!black,line width=0.55pt] (modelT2)--(modelT9);


\foreach \i in {1,...,10}{
  \filldraw[
    fill=layerblue!90!black,
    draw=layerblue!55!black,
    line width=0.2pt
  ]
  (modelB\i) circle (1.55pt);
}

\foreach \i in {1,...,10}{
  \filldraw[
    fill=layerorange!90!black,
    draw=layerorange!55!black,
    line width=0.2pt
  ]
  (modelM\i) circle (1.55pt);
}

\foreach \i in {1,...,10}{
  \filldraw[
    fill=layergreen!90!black,
    draw=layergreen!55!black,
    line width=0.2pt
  ]
  (modelT\i) circle (1.55pt);
}

\node[anchor=east,text=layerblue!75!black]
at (0.30,0.30) {$t$};

\node[anchor=east,text=layerorange!80!black]
at (0.30,1.65) {$t+1$};

\node[anchor=east,text=layergreen!70!black]
at (0.30,3.00) {$t+2$};

\end{scope}


\draw[
  -{Latex[length=3mm]},
  draw=gray!65,
  line width=1.2pt
]
(15.80,9.00)--(15.80,8.45);


\draw[
  rounded corners=3pt,
  draw=orange!65!red,
  fill=orange!5,
  line width=0.9pt
]
(0.20,4.35) rectangle (21.40,8.45);

\node[font=\bfseries\Large]
at (10.80,8.00)
{3. Structure of the latent effect process};


\draw[
  rounded corners=2pt,
  draw=blue!45,
  fill=white,
  line width=0.8pt
]
(0.55,4.70) rectangle (11.00,7.60);

\node[
  font=\bfseries\large,
  text=blue!70!black
]
at (5.38,7.22)
{A. Cross-area dependence};

\fill[black!8]
(2.70,5.72) ellipse (2.10 and 0.72);

\fill[layerblue!10]
(2.70,5.77) ellipse (1.95 and 0.62);

\def\FlatNodes{
1/1.20/5.70,
2/2.00/6.08,
3/2.95/6.18,
4/3.80/6.18,
5/4.35/6.00,
6/1.45/5.25,
7/2.45/5.45,
8/3.20/5.43,
9/3.85/5.28,
10/3.52/5.70
}

\begin{scope}

\clip (0.65,4.80) rectangle (4.80,6.95);

\foreach \id/\xx/\yy in \FlatNodes{
  \coordinate (modelArea\id) at (\xx,\yy);
}

\foreach \u/\v in \YearBlueEdges{
  \draw[layerblue!85!black,line width=0.65pt]
  (modelArea\u)--(modelArea\v);
}

\foreach \i in {1,...,10}{
  \filldraw[
    fill=layerblue!92!black,
    draw=layerblue!60!black,
    line width=0.2pt
  ]
  (modelArea\i) circle (1.75pt);
}

\end{scope}

\node[
  font=\scriptsize,
  text=layerblue!80!black
]
at (2.75,6.70)
{annual graph at year $t$: $\mathcal G_t^{(\mathrm D)}$};

\draw[
  rounded corners=2pt,
  draw=blue!40,
  fill=blue!4,
  line width=0.7pt
]
(4.90,4.98) rectangle (10.85,6.00);

\node[font=\small]
at (7.80,5.50)
{$
\bm\varepsilon_t
\mid\tau_\phi,\lambda
\sim
\mathcal N_A
\left[
\bm0,
\left\{
\tau_\phi
\bm R_t^{(\mathrm D)}(\lambda)
\right\}^{-1}
\right]
$};


\node[
  font=\bfseries,
  text=gray!70!black
]
at (11.80,6.05)
{joint};

\draw[
  -{Latex[length=2.5mm]},
  line width=1pt,
  draw=blue!55
]
(11.45,6.05)--(11.05,6.05);

\draw[
  -{Latex[length=2.5mm]},
  line width=1pt,
  draw=orange!70!red
]
(12.15,6.05)--(12.55,6.05);


\draw[
  rounded corners=2pt,
  draw=orange!65!red,
  fill=white,
  line width=0.8pt
]
(12.55,4.70) rectangle (21.05,7.60);

\node[
  font=\bfseries\large,
  text=orange!85!red
]
at (16.80,7.22)
{B. Temporal evolution};

\node[
  circle,
  minimum size=2.2mm,
  inner sep=0pt,
  draw=layerblue!70!black,
  fill=layerblue!22,
  line width=0.6pt
]
(phi1) at (14.00,5.08) {};

\node[
  circle,
  minimum size=2.2mm,
  inner sep=0pt,
  draw=layerorange!75!black,
  fill=layerorange!22,
  line width=0.6pt
]
(phi2) at (14.00,5.88) {};

\node[
  circle,
  minimum size=2.2mm,
  inner sep=0pt,
  draw=layergreen!70!black,
  fill=layergreen!22,
  line width=0.6pt
]
(phi3) at (14.00,6.68) {};

\draw[myarrow] (phi1.north)--(phi2.south);
\draw[myarrow] (phi2.north)--(phi3.south);

\node[anchor=west,text=layerblue!75!black]
at (14.45,5.08) {$\bm\phi_t$};

\node[anchor=west,text=layerorange!85!black]
at (14.45,5.88) {$\bm\phi_{t+1}$};

\node[anchor=west,text=layergreen!70!black]
at (14.45,6.68) {$\bm\phi_{t+2}$};

\draw[
  rounded corners=2pt,
  draw=orange!50!red,
  fill=orange!4,
  line width=0.7pt
]
(16.20,4.98) rectangle (20.45,6.00);

\node[font=\large]
at (18.33,5.40)
{$
\bm\phi_t
=
\rho\bm\phi_{t-1}
+
\bm\varepsilon_t
$};


\draw[
  rounded corners=3pt,
  draw=purple!45,
  fill=purple!4,
  line width=0.9pt
]
(0.20,0.10) rectangle (10.20,3.55);

\node[font=\bfseries\Large]
at (5.20,3.13)
{4. Joint dependence};

\draw[
  rounded corners=2pt,
  draw=purple!40,
  fill=white,
  line width=0.8pt
]
(0.55,0.50) rectangle (9.85,2.72);

\node[font=\small]
at (5.20,1.95)
{$
\bm Q_\phi^{(\mathrm D)}
=
\tau_\phi
\mathbb B_\rho^\top
\mathcal R^{(\mathrm D)}(\lambda)
\mathbb B_\rho
$};

\node[font=\large]
at (5.20,1.10)
{$
\bm\phi
\mid\tau_\phi,\rho,\lambda
\sim
\mathcal N_{AT}
\left[
\bm0,
\left\{
\bm Q_\phi^{(\mathrm D)}
\right\}^{-1}
\right]
$};


\draw[
  rounded corners=3pt,
  draw=blue!45,
  fill=blue!4,
  line width=0.9pt
]
(11.15,0.10) rectangle (21.40,3.55);

\node[font=\bfseries\Large]
at (16.28,3.13)
{5. Hazard model};

\draw[
  rounded corners=2pt,
  draw=blue!40,
  fill=white,
  line width=0.8pt
]
(11.50,0.50) rectangle (21.05,2.72);

\node[font=\Large]
at (16.28,1.70)
{$
\eta_{is}
=
\beta_0
+
\bm b_s^\top\bm\alpha_f
+
\bm z_i^\top\bm\beta
+
\phi_{a_i,\kappa_i(s)}
$};


\draw[
  -{Latex[length=3mm]},
  draw=gray!65,
  line width=1.2pt
]
(5.20,4.35)--(5.20,3.55);

\draw[
  -{Latex[length=3mm]},
  draw=gray!65,
  line width=1.2pt
]
(10.20,1.82)--(11.15,1.82);

\end{tikzpicture}%
}

\caption{
Framework of the proposed model. For simplicity,
the cross-area dependence structure is illustrated
using the graph from a single year.
}
\label{fig:graph_temporal_model}

\end{figure}
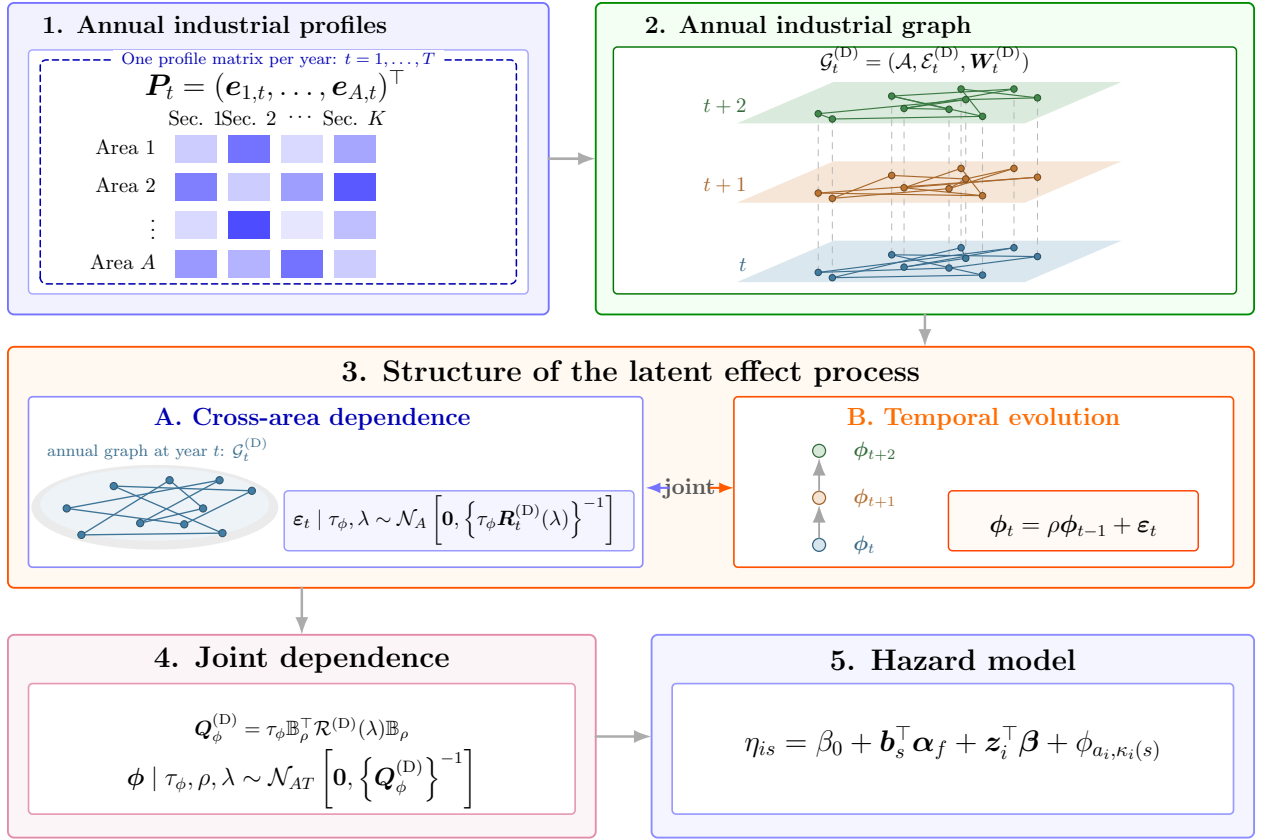

This non-separability is a key feature of the proposed model. Although latent effects evolve through an autoregressive process, their cross-area dependence is structured by industrial graphs that may themselves change as industrial compositions evolve. Therefore, the temporal evolution of latent risk is embedded within an evolving dependence, allowing changes in industrial proximity to shape the joint dependence of latent effects, as illustrated in Fig.~\ref{fig:graph_temporal_model}.

\subsection{Identification}
\label{sec:identification}

The latent effect enters the linear predictor differently
under the static and dynamic specifications. Under the static
specification,
\begin{equation}
\eta_{is}^{(\mathrm S)}
=
\eta_{is}^{(0)}
+
u_{a_i},
\label{eq:static_estimation_predictor}
\end{equation}
where \(u_{a_i}\) is the time-invariant latent effect
associated with area \(a_i\).

Under the dynamic specification,
\begin{equation}
\eta_{is}^{(\mathrm D)}
=
\eta_{is}^{(0)}
+
\phi_{a_i,\kappa_i(s)},
\label{eq:dynamic_estimation_predictor}
\end{equation}
where \(t=\kappa_i(s)\) denotes the year corresponding to duration interval \(s\), and \(\phi_{a_i,t}\) is the latent effect for area \(a_i\) in year \(t\).

To distinguish the latent effects from the overall level by the intercept, we impose sum-to-zero constraints.
For the static specification,
\begin{equation}
\bm 1_A^\top\bm u
=
\sum_{a=1}^{A}u_a
=
0.
\label{eq:static_sum_to_zero}
\end{equation}

For the dynamic, the corresponding constraint is imposed within each year:
\begin{equation}
\bm 1_A^\top\bm\phi_t
=
\sum_{a=1}^{A}\phi_{a,t}
=
0,
\qquad
t=1,\ldots,T.
\label{eq:industry_sum_to_zero}
\end{equation}

Under these constraints, \(\bm u\) represents cross-area deviations from the overall latent effect under the static
specification, whereas \(\bm\phi_t\) represents cross-area deviations from the average latent effect in year \(t\) under
the dynamic specification.

\section{Estimation}
\label{sec:bayesian_estimation}

This study estimates using the Integrated Nested Laplace Approximation (INLA) \citep{rue2009approximate}. Both static and dynamic specifications belong to the class of latent Gaussian models. Conditional on a low-dimensional set of parameters, the coefficients, duration effect, and latent effects are jointly Gaussian, whereas the observed mortgage default indicators follow a non-Gaussian Bernoulli likelihood. The corresponding matrices are sparse: the duration effect inherits sparsity from the spline penalty, while the latent effects inherit sparsity from the cross-area dependence. Under the dynamic specification, the first-order temporal evolution preserves this sparse structure across time. INLA exploits the conditional Gaussian representation and sparsity to approximate posterior marginals without requiring direct integration over the full high-dimensional latent field.

Under the static specification, define $\bm\chi_{\mathrm S}=\left(\beta_0,\bm\beta^\top,\bm\alpha_f^\top,\bm u^\top\right)^\top$, with parameter vector $\bm\vartheta_{\mathrm S}=\left(\tau_f,\tau_u,\lambda\right)^\top$. Under the dynamic specification, define $\bm\chi_{\mathrm D}=\left(\beta_0,\bm\beta^\top,\bm\alpha_f^\top,\bm\phi^\top\right)^\top$, and let $\bm\vartheta_{\mathrm D}=\left(\tau_f,\tau_\phi,\lambda,\rho\right)^\top$. \(\tau_f\) controls the smoothness of the duration effect.

For every loan \(i\) and observed duration interval
\(s=1,\ldots,\widetilde T_i\), let
\(Y_{is}\in\{0,1\}\) denote the default indicator. Conditional on the
latent field,
\begin{equation}
Y_{is}
\mid
\bm\chi_g
\sim
\operatorname{Bernoulli}
\left(
h_{is}^{(g)}
\right),
\qquad
h_{is}^{(g)}
=
\operatorname{logit}^{-1}
\left(
\eta_{is}^{(g)}
\right),
\label{eq:estimation_observation_model}
\end{equation}
where the predictors $\eta_{is}^{(g)}$ are defined in
Section~\ref{sec:identification}.

The corresponding likelihood is
\begin{equation}
\mathcal L_g
\left(
\bm y
\mid
\bm\chi_g
\right)
=
\prod_{i=1}^{N}
\prod_{s=1}^{\widetilde T_i}
\left(
h_{is}^{(g)}
\right)^{y_{is}}
\left(
1-h_{is}^{(g)}
\right)^{1-y_{is}}.
\label{eq:estimation_likelihood}
\end{equation}

Conditional on the model-specific parameters, the latent vector is
assigned a Gaussian prior,
\begin{equation}
\bm\chi_g
\mid
\bm\vartheta_g
\sim
\mathcal N
\left[
\bm0,
\bm Q_{\chi,g}
\left(
\bm\vartheta_g
\right)^{-1}
\right],
\qquad
g\in\{\mathrm S,\mathrm D\},
\label{eq:latent_gaussian_prior}
\end{equation}
subject to the constraints described in Section~\ref{sec:identification}.

Under the static specification, the block-diagonal precision matrix is
\begin{equation}
\bm Q_{\chi,\mathrm S}
\left(
\bm\vartheta_{\mathrm S}
\right)
=
\operatorname{blockdiag}
\left\{
\sigma_0^{-2},
\;
\sigma_\beta^{-2}I,
\;
\tau_f\bm Q_f,
\;
\bm Q_u^{(\mathrm S)}
\right\},
\label{eq:static_full_precision}
\end{equation}
where $\bm Q_u^{(\mathrm S)}=\tau_u
\bm R_{t_0}^{(\mathrm S)}(\lambda)$.

Under the dynamic specification,
\begin{equation}
\bm Q_{\chi,\mathrm D}
\left(
\bm\vartheta_{\mathrm D}
\right)
=
\operatorname{blockdiag}
\left\{
\sigma_0^{-2},
\;
\sigma_\beta^{-2}I,
\;
\tau_f\bm Q_f,
\;
\bm Q_\phi^{(\mathrm D)}
\right\},
\label{eq:dynamic_full_precision}
\end{equation}
where $\bm Q_\phi^{(\mathrm D)}=\tau_\phi\mathbb B_\rho^\top\mathcal R^{(\mathrm D)}(\lambda)\mathbb B_\rho$.

The quantities \(\sigma_0^2\) and \(\sigma_\beta^2\) are fixed at large values, corresponding to weakly informative Gaussian priors for the intercept and regression coefficients. The block \(\tau_f\bm Q_f\) is the precision associated with the duration-spline coefficients, where \(\bm Q_f\) determines the spline penalty. 

We assume prior independence among the parameters within each specification. Under the static specification,
\begin{equation}
p
\left(
\bm\vartheta_{\mathrm S}
\right)
=
p(\tau_f)\,
p(\tau_u)\,
p(\lambda),
\label{eq:static_hyperprior}
\end{equation}
whereas under the dynamic specification,
\begin{equation}
p
\left(
\bm\vartheta_{\mathrm D}
\right)
=
p(\tau_f)\,
p(\tau_\phi)\,
p(\lambda)\,
p(\rho).
\label{eq:dynamic_hyperprior}
\end{equation}

Parameters common to both specifications are assigned the same prior. Besides, the parameters \(\tau_u\) and \(\tau_\phi\) are also calibrated using the same scaling, as in Section~\ref {sec:graph_scaling}.

Combining the Bernoulli likelihood, the conditional latent Gaussian prior, and the model-specific parameter prior gives
\begin{equation}
p
\left(
\bm\chi_g,
\bm\vartheta_g
\mid
\bm y
\right)
\propto
\mathcal L_g
\left(
\bm y
\mid
\bm\chi_g
\right)
\,
p
\left(
\bm\chi_g
\mid
\bm\vartheta_g
\right)
\,
p
\left(
\bm\vartheta_g
\right),
\qquad
g\in\{\mathrm S,\mathrm D\}.
\label{eq:joint_posterior}
\end{equation}

Eq.\eqref{eq:joint_posterior} makes explicit how the three layers of the model interact: the observed mortgage histories contribute through the Bernoulli likelihood \(\mathcal L_g\left(\bm y\mid\bm\chi_g\right)\), the duration effect and the latent effects enter through the latent Gaussian prior \(p\left(\bm\chi_g\mid\bm\vartheta_g\right)\), and uncertainty in parameters enters through the prior \(p\left(\bm\vartheta_g\right)\).

Direct evaluation of the joint posterior is computationally demanding, particularly under the dynamic specification, which contains \(AT\) latent effects and the observation likelihood is non-Gaussian. INLA avoids direct integration over the
full latent field by exploiting the low dimension of \(\bm\vartheta_g\) and the sparsity of \(\bm Q_{\chi,g}(\bm\vartheta_g)\).

Integrating out the latent Gaussian field gives the parameter posterior
\begin{equation}
p
\left(
\bm\vartheta_g
\mid
\bm y
\right)
\propto
p
\left(
\bm\vartheta_g
\right)
\int
\mathcal L_g
\left(
\bm y
\mid
\bm\chi_g
\right)
\,
p
\left(
\bm\chi_g
\mid
\bm\vartheta_g
\right)
\,d\bm\chi_g.
\label{eq:hyperparameter_posterior}
\end{equation}
The integral in
Eq.\eqref{eq:hyperparameter_posterior}
does not have a closed-form solution since the Bernoulli observation
model is not conjugate to the Gaussian latent field.

For a fixed value of \(\bm\vartheta_g\), the conditional posterior of
the latent field is
\begin{equation}
p
\left(
\bm\chi_g
\mid
\bm\vartheta_g,
\bm y
\right)
\propto
\mathcal L_g
\left(
\bm y
\mid
\bm\chi_g
\right)
\,
p
\left(
\bm\chi_g
\mid
\bm\vartheta_g
\right).
\label{eq:conditional_latent_posterior}
\end{equation}

Let \(\bm\chi_g^\ast(\bm\vartheta_g)\) denote the mode of Eq.\eqref{eq:conditional_latent_posterior}. For the Bernoulli-logit likelihood, this mode can be obtained using Newton-type iterations that exploit the sparse precision structure
\citep{rue2009approximate}. Let $\widetilde p_G\left(\bm\chi_g\mid\bm\vartheta_g,\bm y\right)$ denote the Gaussian approximation to the conditional latent posterior obtained from a second-order expansion of its log-density around \(\bm\chi_g^\ast(\bm\vartheta_g)\). The nested Laplace approximation to the parameter posterior is then
\begin{equation}
\widetilde p
\left(
\bm\vartheta_g
\mid
\bm y
\right)
\propto
\left.
\frac{
\mathcal L_g
\left(
\bm y
\mid
\bm\chi_g
\right)
\,
p
\left(
\bm\chi_g
\mid
\bm\vartheta_g
\right)
\,
p
\left(
\bm\vartheta_g
\right)
}{
\widetilde p_G
\left(
\bm\chi_g
\mid
\bm\vartheta_g,
\bm y
\right)
}
\right|_{
\bm\chi_g
=
\bm\chi_g^\ast(\bm\vartheta_g)
}.
\label{eq:inla_hyperparameter_approximation}
\end{equation}

Once the parameter posterior has been approximated, the marginal posterior of a generic latent component \(\chi_{g,j}\) is obtained from
\begin{equation}
p
\left(
\chi_{g,j}
\mid
\bm y
\right)
=
\int
p
\left(
\chi_{g,j}
\mid
\bm\vartheta_g,
\bm y
\right)
p
\left(
\bm\vartheta_g
\mid
\bm y
\right)
\,d\bm\vartheta_g.
\label{eq:latent_marginal_posterior}
\end{equation}

The prior setting of this study is presented in Appendix~\ref{si_sec:prior}.

\section{Evaluation}
\label{sec:evaluation}
The empirical evaluation is designed around two questions. First, the dynamic specification is meaningful only if industrial composition changes materially over time, and any such change translates into variation in the graph constructed from it. Second, we assess whether the dynamic specification improves the discrimination of default risk relative to the fixed-graph benchmark.

\subsection{Industrial Information Diagnosis}

Before fitting the model, we first examine whether the underlying industrial composition, and hence the graph, change over time. A given degree of compositional drift may in principle leave the relative similarity largely unchanged, or it may reshape that similarity substantially.

\paragraph{Industrial composition.} We characterise each area's industrial composition by the distribution of employment across industrial sectors, following \citet{ellison1997geographic,howard2016measuring,gervais2024regional}. Recall from Eq.\eqref{eq:industry_profile} that \(\bm e_{a,t}\in\Delta^{K-1}\) denotes the employment composition of area \(a\) across \(K\) industrial sectors in year \(t\). We measure the year-to-year change in an area's industrial composition using the total variation distance between consecutive annual compositions,
\begin{equation}
\mathrm{TV}_{a,t}
=
\frac12
\sum_{k=1}^{K}
\left|
e_{ak t}-e_{ak,t-1}
\right|,
\qquad
a\in\mathcal A,
\quad
t=2,\ldots,T.
\label{eq:area_profile_tv_distance}
\end{equation}
Since \(\bm e_{a,t},\bm e_{a,t-1}\in\Delta^{K-1}\) are both probability vectors, \(\mathrm{TV}_{a,t}\in[0,1]\), with \(\mathrm{TV}_{a,t}=0\) if and only if area \(a\)'s industrial composition is unchanged between years \(t-1\) and \(t\), and \(\mathrm{TV}_{a,t}=1\) only in the extreme case in which the two compositions have entirely disjoint industry support. We summarise the magnitude of compositional change across areas in year \(t\) by
\begin{equation}
\overline{\mathrm{TV}}_t
=
\frac{1}{A}
\sum_{a=1}^{A}
\mathrm{TV}_{a,t},
\qquad
t=2,\ldots,T,
\label{eq:average_profile_tv_distance}
\end{equation}
the net drift of each area's composition relative to its starting point
\begin{equation}
\mathrm{TV}_{a,t}^{\mathrm{net}}
=
\frac12
\sum_{k=1}^{K}
\left|
e_{akt}-e_{ak,t_0}
\right|,
\qquad
a\in\mathcal A,
\quad
t=2,\ldots,T.
\label{eq:cumulative_profile_tv_distance}
\end{equation}

and the average net drift across areas in year $t$ by
\begin{equation}
\overline{\mathrm{TV}_{t}^{\mathrm{net}}}
=
\frac{1}{A}
\sum_{a=1}^{A}
\mathrm{TV}_{a,t}^{\mathrm{net}},
\qquad
t=2,\ldots,T,
\label{eq:avergage_cumulative_profile_tv_distance}
\end{equation}

These measures distinguish short-term fluctuations in industrial composition from more persistent departures from the baseline. A relatively large $\overline{\mathrm{TV}}_t$ accompanied by a small $\overline{\mathrm{TV}}_t^{\mathrm{net}}$ suggests that regional industrial compositions undergo noticeable annual adjustments without necessarily moving far from their initial compositions. Conversely, an increasing $\overline{\mathrm{TV}}_t^{\mathrm{net}}$ indicates that industrial compositions are becoming different from those observed in the baseline year. Such changes may have implications for the industrial proximity used to construct the graph. However, compositional change does not translate mechanically into graph evolution, since neighbourhoods and edge weights depend on relative similarities between areas.

\paragraph{Industrial graph.} 

We then characterise the annual dynamic graph to assess whether the compositional change translates into a corresponding reconfiguration of the industry similarity. For year \(t\), recall the edge set $\mathcal E_t$, changes
in graph topology between consecutive years are summarised by the
Jaccard distance
\begin{equation}
\mathcal J_t
=
1-
\frac{
|\mathcal E_t\cap \mathcal E_{t-1}|
}{
|\mathcal E_t\cup \mathcal E_{t-1}|
},
\qquad
t=2,\ldots,T.
\label{eq:annual_jaccard_distance}
\end{equation}
A value of \(\mathcal J_t\) close to zero indicates that consecutive annual graphs share nearly identical edge sets, suggesting that neighbourhoods remain largely unchanged. Larger values, by contrast, indicate greater changes in graph topology through the formation and disappearance of edges.

\subsection{Model Performance}
\label{sec:model_performance}

We evaluate model performance along three complementary dimensions, including in-sample fit, in-sample predictive adequacy, and out-of-sample discrimination.

\paragraph{Model fitting}
This study reports the Deviance Information Criterion (DIC) \citep{spiegelhalter2002bayesian} and the Watanabe Akaike Information Criterion (WAIC) \citep{watanabe2010asymptotic}, with lower values indicating a more favourable balance between model fit and effective complexity. However, these criteria are used for assessing model fit; they do not assess predictive performance. 

\paragraph{Predictive adequacy}
Predictive adequacy is further assessed through conditional predictive ordinates. Let \(j=1,\ldots,n\) index the loan--month observation, and let \(\bm y_{-j}\) denote all observations except \(y_j\). The conditional predictive ordinate is defined as $\mathrm{CPO}_j
=
p(y_j\mid\bm y_{-j})$, and represents the leave-one-observation-out predictive density assigned to the observation \(y_j\) \citep{geisser1979predictive}.

The individual predictive ordinates are summarised by the log pseudo-marginal likelihood \citep{gelfand1992model},
\begin{equation}
\mathrm{LPML}
=
\sum_{j=1}^{n}
\log \mathrm{CPO}_j.
\label{eq:lpml}
\end{equation}
Larger values of LPML indicate better local predictive adequacy.

\paragraph{Discrimination}
Although the above provide measures of model fit and predictive adequacy, they remain within-sample assessments. We therefore evaluate out-of-sample discrimination on a non-overlapping origination cohort by using the area under the receiver operating characteristic curve (AUC) \citep{hanley1982meaning} and the Kolmogorov-Smirnov (KS) statistic \citep{smirnov1948table}. This design assesses whether the discrimination of the proposed model could be generalised to mortgages originated in subsequent vintages.

\section{Empirical analysis}
\label{sec:results}
\subsection{Data}
\label{sec:data}

The empirical analysis relies on two primary data sources. The first is the \emph{Freddie Mac Single-Family Loan-Level Dataset}, a publicly available repository containing loan-level and borrower-level characteristics and monthly performance histories for conventional mortgages purchased or guaranteed by Freddie Mac. The second is the U.S. Census Bureau's County Business Patterns (CBP), from which we construct annual industrial composition and the graphs.

The mortgage sample consists of loans originated from January 2007 to December 2008 and followed through their subsequent monthly performance histories until December 2020. The outcome of interest is mortgage default, which we define as the first occurrence of serious delinquency, corresponding to a delinquency status of 90 days or more past due. The sample contains 91{,}563 loans and 4{,}151{,}689 loan-month observations, of which 6{,}640 loans experience a default over the observation period. Each loan contributes one observation for every month in which it remains at risk, including the month of default, and exits the risk set thereafter. Besides, we use the county as the primary areal unit. The covariates used in this paper are presented in Table~\ref{tab:covariate_definitions}.

\begin{table}[H]
\footnotesize
\caption{Explanatory variables}
\label{tab:covariate_definitions}
\centering
\begin{tabular}{p{4cm}p{2.0cm}p{9cm}}
\toprule
Explanatory variable & Type & Description \\
\midrule
\multicolumn{3}{l}{\textit{Loan Information}} \\
Original UPB          & Numeric     & Unpaid principal balance at origination \\
Original Interest Rate & Numeric    & Note rate at origination \\
Original Loan Term    & Numeric     & Original number of scheduled monthly mortgage payments \\
DTI                    & Numeric     & Original debt-to-income ratio \\
Loan Purpose           & Categorical & Purchase, cash-out refinance, or no-cash-out refinance \\
\midrule
\multicolumn{3}{l}{\textit{Borrower Information}} \\
Credit Score            & Numeric     & Credit score reported at loan origination \\
Number of Borrowers     & Categorical     & Number of borrowers obligated on the loan \\
\midrule
\multicolumn{3}{l}{\textit{Property Information}} \\
LTV                     & Numeric     & Original loan-to-value ratio \\
Property Type           & Categorical & Single-family, planned unit development, condominium, manufactured housing, or co-op \\
Occupancy Status        & Categorical & Primary residence, investment property, or second home \\
Number of Units         & Numeric     & Number of dwelling units securing the loan \\
\bottomrule
\end{tabular}
\end{table}

\subsection{Models}

We consider three specifications, as summarised in Table~\ref{tab:model_specs}. All share the same default definition and risk sets, duration effects, borrower and loan covariates, prior families, and evaluation; they differ only in the specification of the latent dependence.

For \(M_1\), the model omits residual latent dependence and serves as the baseline.

For \(M_2\) and \(M_3\), we construct a nearest-neighbour graph with five neighbours, following \citet{calabrese2022modelling}, but define proximity in terms of industrial composition. Specifically, for each county-year, we use CBP employment data to compute the share of employment in each sector. These shares constitute the county's industrial profile, as defined in Eq.\eqref{eq:industry_profile}. Before computing industry distances, each of the employment shares is standardised annually using its contemporaneous cross-county mean and standard deviation across the 679 counties. We then calculate Euclidean distance between the resulting standardised profiles and select, for each county, the five nearest industry neighbours. 
Since the resulting relationship need not be symmetric, we symmetrise the relation before constructing the undirected graph. An edge is therefore present whenever either county includes the other among its five nearest neighbours.
For the static specification \(M_2\), we construct the graph using industrial composition observed in 2007, the first year of the estimation sample, and hold the graph fixed throughout estimation and evaluation. We do not add temporal dependence to $M_2$, so that $M_2$ remains the static benchmark against which the dynamic specification is assessed. For \(M_3\), graphs are reconstructed annually, based on the CBP employment shares in each year. Importantly, each annual graph is connected, with a single connected component in every year, which satisfies the assumptions of Section~\ref{sec:graph_scaling}, Lemma~\ref{lem:laplacian_psd} and Proposition~\ref{prop:graph_scaling} in Supplementary Material~\ref{app:proofs}.

\begin{table}[H]
\footnotesize
  \centering
  \caption{Model design.}
  \label{tab:model_specs}
  \begin{tabular}{@{} l p{4.5cm} p{4.2cm} l @{}}
    \toprule
    Model & Geographic term & Industrial term & Linear predictor \\
    \midrule
    \(M_1\)
    & None
    & None
    & \(\eta_{is}^{(0)}\) \\[6pt]

    \(M_2\)
    & None
    & Static industrial graph, \newline time-invariant effect
    & \(\eta_{is}^{(0)} + u_{a_i}\) \\[6pt]

    \(M_3\)
    & None
    & Dynamic graph, \newline time-varying effect
    & \(\eta_{is}^{(0)} + \phi_{a_i,\kappa_i(s)}\) \\
    \bottomrule
  \end{tabular}
\end{table}

\subsection{Results}
\subsubsection{Industrial Information Diagnosis Results}
\label{sec:diagnosis}

Before estimating, we examine whether the underlying industrial composition has sufficient temporal variation to allow the industrial graph to evolve over time, and whether this compositional change translates into a corresponding reconfiguration of the graph.

\begin{figure}[H]
    \centering
    \includegraphics[width=0.8\linewidth]{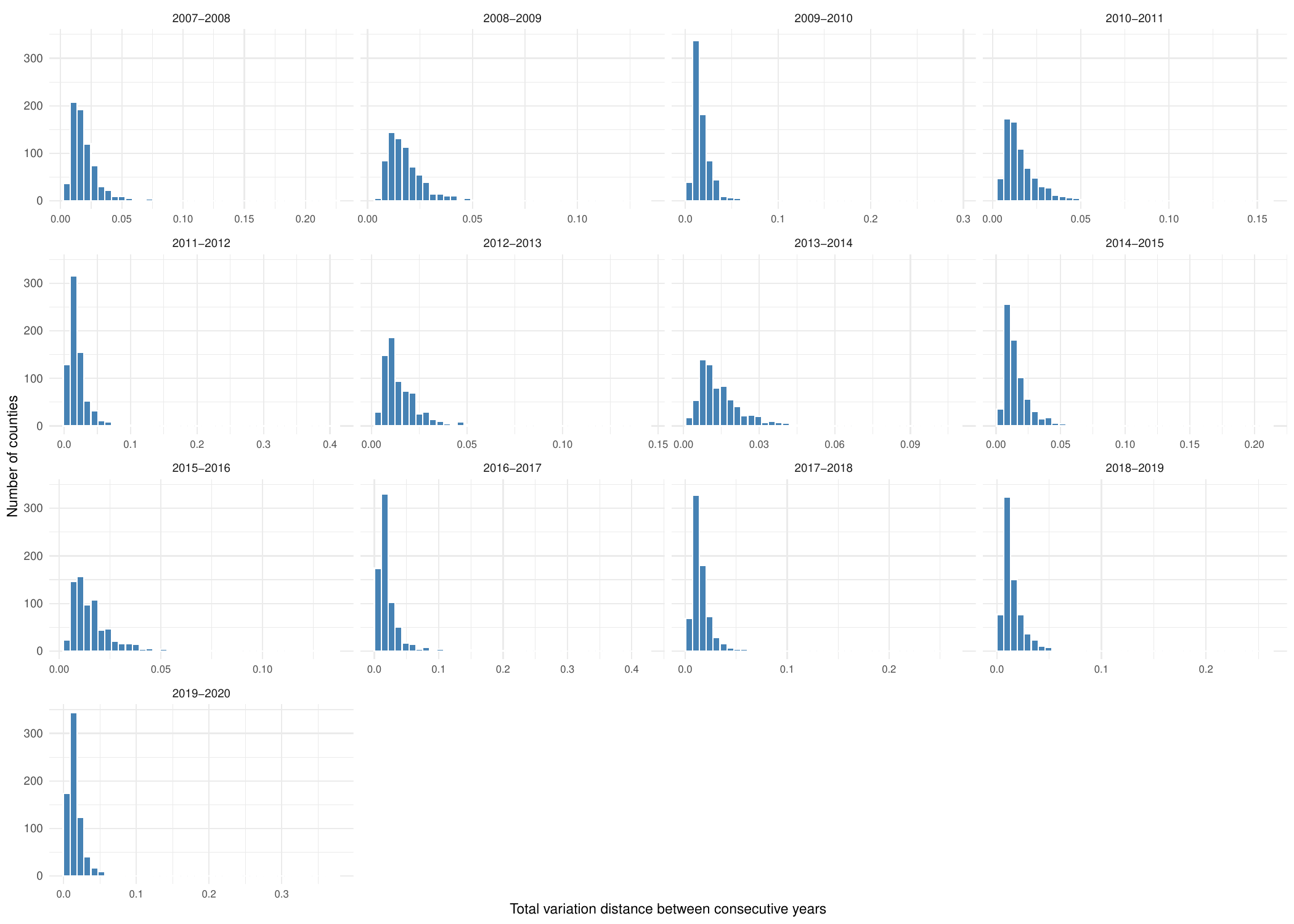}
    \caption{Total variation in county-level industry
composition from 2007 to 2020.}
    \label{fig:tv_distance_by_year}
\end{figure}

\paragraph{Industrial composition.}
Fig.~\ref{fig:tv_distance_by_year} shows the full distribution of $\mathrm{TV}_{a,t}$ across counties for each of the 13 year-to-year transitions. The distributions are unimodal and right-skewed in every transition, with the large majority of counties clustered at low values of $\mathrm{TV}_{a,t}$. Table~\ref{tab:tv_jaccard_cumulative} averages across areas and reports the average total variation over 2007-2020. The compositional change is modest yet persistent, ranging from 0.0155 to 0.0222. Importantly, these movements do not cancel out over the estimation period.

\begin{figure}[H]
    \centering
    \includegraphics[width=0.8\linewidth]{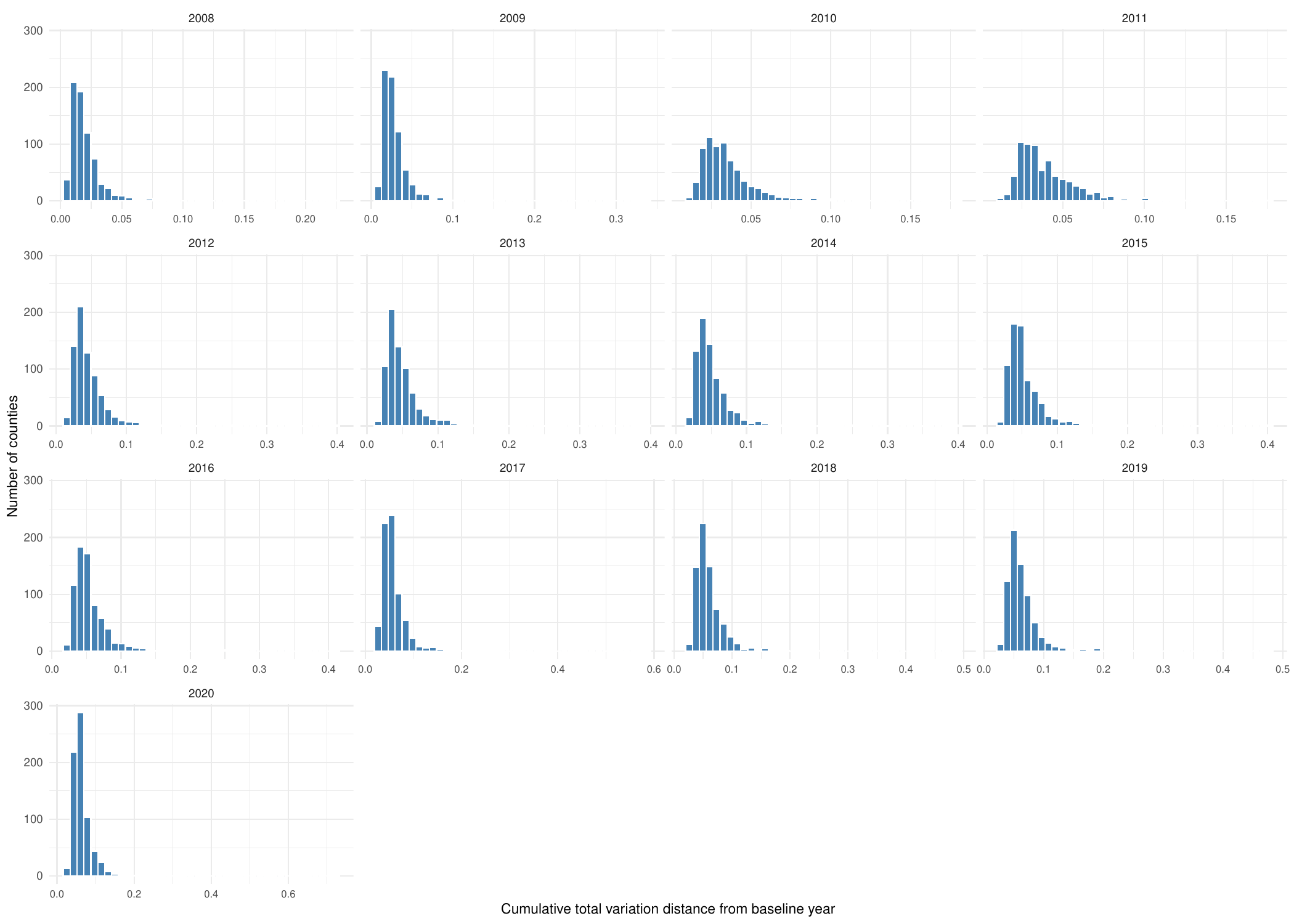}
    \caption{Net total variation drift in county-level industry
composition relative to the 2007 baseline, by year.}
    \label{fig:tvcum_by_year}
\end{figure}

Fig.~\ref{fig:tvcum_by_year} shows the corresponding distribution of the net drift $\mathrm{TV}^{\mathrm{net}}_{a,t}$ across the areas for each year. The distributions shift steadily to the right over time. Table~\ref{tab:tv_jaccard_cumulative} tracks the average $\overline{\mathrm{TV}^{\mathrm{net}}_t}$ every year, also revealing that this net drift grows steadily and monotonically throughout the estimation period. This monotonic growth confirms that the modest yearly compositional movements accumulate into sustained structural reallocation; a county's industrial composition in 2020 is, on average, meaningfully different from its 2007 baseline.

\paragraph{Industrial graph.}
Even though the compositional movement is modest, the changes in the industrial graph are substantial. Across the 13 annual transitions, the average Jaccard distance is 0.438; that is, on average only 56.2\% of the edges present in either of two consecutive years are present in both. This marked instability arises since even a modest change in an area's industrial composition can change which other areas emerge as its neighbours, even though the underlying compositional change appears small in absolute terms. Therefore, a graph calibrated only at a fixed year would misrepresent the similarity for much of the sample period.

\begin{table}[H]
\footnotesize
  \centering
  \caption{Total variation, graph Jaccard distance, and net
  total variation drift over time.}
  \label{tab:tv_jaccard_cumulative}
  \begin{tabular}{@{} l S[table-format=1.4] S[table-format=1.4]
                  @{\hspace{2.5em}}
                  l S[table-format=1.4] S[table-format=1.4] @{}}
    \toprule
    Transition & {$\overline{\mathrm{TV}}_t$} & {$\mathcal{J}_t$}
               & Transition & {$\overline{\mathrm{TV}}_t$} & {$\mathcal{J}_t$} \\
    \midrule
    2007--2008 & 0.0195 & 0.4787 & 2014--2015 & 0.0164 & 0.4412 \\
    2008--2009 & 0.0188 & 0.4333 & 2015--2016 & 0.0166 & 0.4254 \\
    2009--2010 & 0.0179 & 0.4350 & 2016--2017 & 0.0222 & 0.4869 \\
    2010--2011 & 0.0167 & 0.4407 & 2017--2018 & 0.0160 & 0.4028 \\
    2011--2012 & 0.0207 & 0.4853 & 2018--2019 & 0.0158 & 0.4153 \\
    2012--2013 & 0.0162 & 0.4156 & 2019--2020 & 0.0163 & 0.4130 \\
    2013--2014 & 0.0155 & 0.4159 &            &        &        \\
    \midrule
    \multicolumn{6}{@{}l}{\textit{Net total variation drift relative
    to the 2007 baseline}} \\
    \midrule
    Year & { $\overline{\mathrm{TV}^{\mathrm{net}}_t}$} & {}
         & Year & {$\overline{\mathrm{TV}^{\mathrm{net}}_t}$} & {} \\
    \midrule
    2008 & 0.0195 & & 2015 & 0.0538 & \\
    2009 & 0.0286 & & 2016 & 0.0557 & \\
    2010 & 0.0351 & & 2017 & 0.0600 & \\
    2011 & 0.0405 & & 2018 & 0.0619 & \\
    2012 & 0.0461 & & 2019 & 0.0640 & \\
    2013 & 0.0497 & & 2020 & 0.0667 & \\
    2014 & 0.0519 & &      &        & \\
    \bottomrule
  \end{tabular}

\end{table}

\subsubsection{Parameter Estimates}
\label{sec:parameter_estimates}

\begin{figure}[H]
  \centering
  \includegraphics[width=0.6\textwidth]{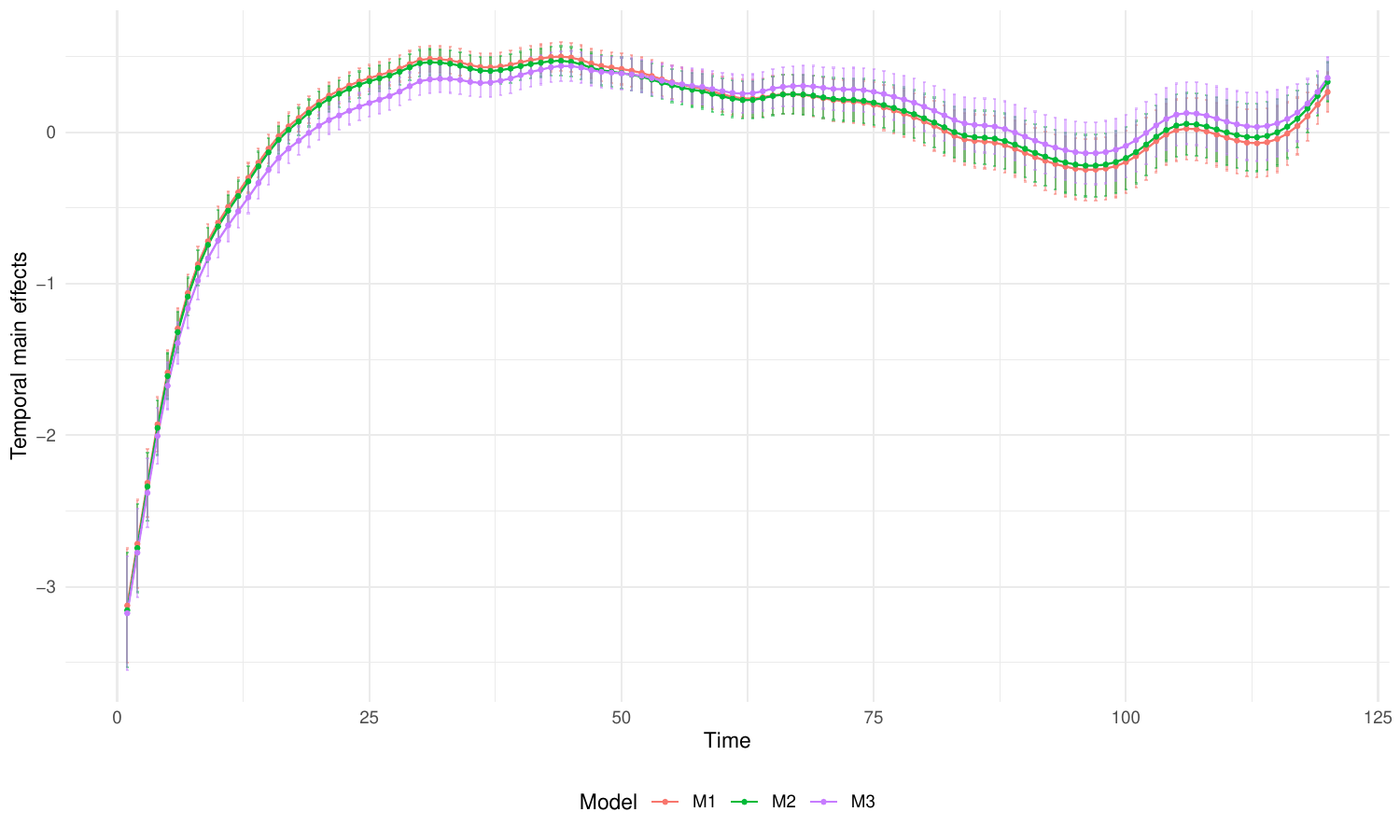}
  \caption{Temporal main effects estimated by models. Error
  bars represent the estimated 95\% credible intervals.}
  \label{fig:temporal_main_effects}
\end{figure}

Table~\ref{tab:parameter_estimates} shows the parameter estimates for the three specifications. We observe that several covariates are estimated consistently across these specifications. The credible intervals of the coefficients of DTI, interest rate, credit score, and number of borrowers almost entirely overlap. The five covariates whose 95\% credible intervals already include zero under $M_1$, loan purpose (N), occupancy status (P), and number of units, remain inconclusive under every specification. However, the coefficient on LTV under $M_1$ is 0.389; once a latent effect is introduced, it rises to 0.452 under $M_2$, and 0.448 under $M_3$. The coefficient on original UPB moves in the opposite direction, falling from 0.237 under $M_1$ to 0.186--0.191 under $M_2$ and $M_3$. In both cases, the sign and statistical significance are preserved across all specifications; the shift instead reflects the latent effect absorbing part of the heterogeneity in default risk that is correlated with local leverage and loan size. Regarding the parameters, the precision of the duration-spline effect, $\tau_f$, is estimated consistently across all specifications, with heavily overlapping credible intervals, as in Fig.~\ref{fig:temporal_main_effects}. $M_3$ shows some differences, in particular for the beginning periods, but overall, the effect of the three models is fairly comparable.

\begin{table}[H]
\footnotesize
\centering
\caption{Parameter estimates for models.}
\label{tab:parameter_estimates}
\resizebox{\textwidth}{!}{%
\begin{tabular}{lccccccccc}
\toprule
 & \multicolumn{3}{c}{$M_1$} & \multicolumn{3}{c}{$M_2$} & \multicolumn{3}{c}{$M_3$} \\
\cmidrule(lr){2-4} \cmidrule(lr){5-7} \cmidrule(lr){8-10}
 & Mean & 2.5\% & 97.5\% & Mean & 2.5\% & 97.5\% & Mean & 2.5\% & 97.5\% \\
\midrule
Intercept           & --5.880 & --5.996 & --5.763 & --6.202 & --6.330 & --6.074 & --6.194 & --6.322 & --6.066 \\
DTI                 & 0.178   & 0.152   & 0.204   & 0.174   & 0.148   & 0.200   & 0.174   & 0.148   & 0.200   \\
Original loan term  & 0.224   & 0.184   & 0.264   & 0.191   & 0.151   & 0.232   & 0.193   & 0.153   & 0.234   \\
Interest rate       & 0.316   & 0.287   & 0.345   & 0.308   & 0.278   & 0.337   & 0.308   & 0.278   & 0.337   \\
LTV                 & 0.389   & 0.353   & 0.424   & 0.452   & 0.414   & 0.490   & 0.448   & 0.411   & 0.486   \\
Number of borrowers & --0.264 & --0.290 & --0.239 & --0.267 & --0.293 & --0.240 & --0.266 & --0.292 & --0.239 \\
Original UPB        & 0.237   & 0.210   & 0.263   & 0.191   & 0.160   & 0.221   & 0.186   & 0.155   & 0.217   \\
Number of units     & --0.010 & --0.035 & 0.015   & 0.008   & --0.016 & 0.033   & 0.009   & --0.016 & 0.034   \\
Credit score        & --0.388 & --0.413 & --0.363 & --0.409 & --0.434 & --0.383 & --0.407 & --0.433 & --0.382 \\
Loan purpose: N     & --0.059 & --0.123 & 0.005   & 0.005   & --0.061 & 0.071   & 0.011   & --0.054 & 0.077   \\
Loan purpose: P     & --0.565 & --0.629 & --0.502 & --0.508 & --0.572 & --0.444 & --0.498 & --0.562 & --0.433 \\
Occupancy: P        & --0.049 & --0.134 & 0.037   & --0.019 & --0.106 & 0.068   & --0.015 & --0.102 & 0.072   \\
Occupancy: S        & 0.080   & --0.053 & 0.214   & --0.030 & --0.167 & 0.107   & --0.013 & --0.150 & 0.124   \\
Property type: CP   & --0.531 & --1.043 & --0.019 & --0.288 & --0.833 & 0.256   & --0.283 & --0.826 & 0.260   \\
Property type: MH   & 0.156   & --0.064 & 0.377   & 0.092   & --0.136 & 0.320   & 0.093   & --0.134 & 0.321   \\
Property type: PU   & --0.125 & --0.223 & --0.026 & --0.089 & --0.195 & 0.017   & --0.094 & --0.200 & 0.012   \\
Property type: SF   & --0.116 & --0.198 & --0.035 & 0.006   & --0.081 & 0.093   & 0.007   & --0.080 & 0.094   \\
\midrule
$\tau_f$            & 1207.7  & 269.7   & 3599.1  & 1198.1  & 267.5   & 3563.1  & 1253.5  & 282.4   & 3702.8  \\
$\tau_u$            &         &         &         & 7.484   & 5.879   & 9.380   &         &         &         \\
$\tau_\phi$         &         &         &         &         &         &         & 18.053  & 12.321  & 25.977  \\
$\rho$              &         &         &         &         &         &         & 0.847   & 0.794   & 0.892   \\
$\lambda$           &         &         &         & 0.971   & 0.917   & 0.994   & 0.985   & 0.959   & 0.997   \\
\bottomrule
\end{tabular}%
}
\end{table}

At the same time, the static specification $M_2$ yields an estimated precision of $\tau_u=7.484$ for the time-invariant latent effect, together with a high structure weight of $\lambda=0.971$. This indicates that the industrial graph captures an important cross-area dependence in mortgage default risk, even when the industrial relationships are held fixed over time. However, the static specification assumes that the latent effect remains unchanged throughout the observation
period, and therefore cannot accommodate changes in the
underlying dependence as industrial compositions evolve.

By contrast, the dynamic specification $M_3$ estimates an innovation precision of $\tau_\phi=18.053$ and an autoregressive coefficient of $\rho=0.847$. The relatively high persistence, while remaining clearly below one, indicates that the latent effect retains a substantial influence from the preceding year without being restricted to remain unchanged. This is the advantage of the dynamic specification over the static alternative: it preserves the persistence of latent mortgage risk while allowing new innovations, structured by the updated industrial graph, to alter the cross-area dependence over time. Besides, the estimated structure weight in $M_3$ is even closer to 1, with a posterior mean of $\lambda=0.985$, indicating that the graph-induced cross-area dependence receives substantially greater weight than the unstructured component. These results provide empirical support for the premise underlying the dynamic industrial graph: areas with similar industrial compositions exhibit residual dependence in mortgage default risk, and the structure governing this dependence can evolve as industrial compositions change. The graph construction described in Section~\ref{sec:graph_induced_dependence} therefore
captures a dependence structure that is strongly supported by the data.

\subsubsection{Model performance}
\label{sec:oot_performance}

We first evaluate the three specifications in terms of in-sample fit and predictive adequacy. As reported in Table~\ref{tab:oot_fit_discrimination}, the dynamic specification $M_3$ consistently outperforms the baseline and static alternatives, attaining the lowest DIC and WAIC and the highest LPML. Relative to the static specification $M_2$, DIC and WAIC decrease by 175.9 and 179.3, respectively, indicating that allowing the graph and latent effects to evolve improves model fit even after accounting for effective model complexity.

We further assess out-of-sample discrimination using a non-overlapping origination cohort comprising mortgages originated between January 2009 and December 2010. This evaluation is particularly relevant since it tests whether the estimated relationships extend beyond the mortgage vintages used for estimation. The dynamic specification again achieves the strongest performance, with an AUC of 0.7816 and a KS statistic of 0.4260. Although the increase compared to the static specification in AUC may appear modest in absolute terms, the economic relevance of this improvement becomes clearer when mortgage risk predictions are translated into lending and portfolio management decisions. Financial institutions typically allocate monitoring resources, determine underwriting thresholds, and prioritise risk interventions across large portfolios, where even relatively small improvements in risk ranking can affect which exposures receive attention. Better identification of high-risk mortgages also therefore influences expected credit losses through the outstanding exposure and potential loss severity associated with the loans identified \citep{verbraken2014development}. 

\begin{table}[H]
\footnotesize
\centering
\caption{In-sample fit, predictive adequacy, and out-of-sample discrimination.}
\label{tab:oot_fit_discrimination}
\begin{tabular}{lrrrrr}
\toprule
Model & DIC & WAIC & LPML & AUC & KS \\
\midrule
$M_1$ & 83{,}044.8 & 83{,}044.5 & $-41{,}522.3$ & 0.7634 & 0.3963 \\
$M_2$     & 82{,}087.6 & 82{,}083.4 & $-41{,}041.7$ & 0.7688 & 0.4048 \\
$M_3$   & 81{,}911.7 & 81{,}904.1 & $-40{,}952.0$ & 0.7816 & 0.4260 \\
\bottomrule
\end{tabular}
\end{table}

\section{Conclusion}
\label{sec:conclusion}
This paper develops a discrete-time survival model that incorporates a dynamic industrial graph to capture the joint evolution of latent risk and cross-area dependence. The motivation is that geographic proximity captures only one dimension of regional dependence: areas that are geographically distant may nevertheless be exposed to similar mortgage risk when they share comparable industrial compositions. Industrial similarity therefore provides an alternative basis for identifying economically connected areas. However, these relationships are not necessarily stable, as changes in local industrial composition may cause previously similar areas to diverge and previously dissimilar areas to become more closely connected. A model based on a fixed graph cannot capture such changes, even if the associated latent effects are allowed to vary over time, since the set of connections through which dependence is represented remains unchanged. Therefore, it may continue to impose dependence between areas that are no longer economically similar while failing to recognise newly emerging relationships. 

We address this limitation by reconstructing the industrial graph annually from contemporaneous industrial composition, allowing both its topology and edge weights to evolve over time. The resulting sequence of graphs determines the cross-area dependence of annual latent effects, while a first-order autoregressive process links the latent effects across successive years. We jointly combine these two components within a proper conditional autoregressive framework, with explicit conditions ensuring the validity and identifiability of the proposed model. Estimation is performed using the Integrated Nested Laplace Approximation, which exploits the sparsity of the joint precision matrix to accommodate a large panel of mortgage observations. We evaluate the proposed model using U.S. mortgage data, comparing a baseline specification with models based on static and dynamic industrial graphs.

The empirical analysis yields four main findings. First, industrial similarity provides a meaningful basis for representing cross-area dependence in mortgage default risk. The high estimated structure weights in both the static and dynamic industrial specifications indicate that graph-induced cross-area dependence plays an important role in explaining residual variation beyond observed borrower and loan characteristics. Second, the estimated persistent value further indicates substantial temporal persistence in latent risk, while allowing the effects of annually updated industrial graphs to enter through new innovations. Third, allowing the graph to evolve improves model performance beyond what can be achieved with a fixed graph. The dynamic specification achieves the lowest DIC and WAIC and the highest LPML, suggesting that changes in industrial relationships contain information about mortgage default risk that is not fully captured by static connections. Fourth, the improvement extends beyond in-sample fit to out-of-sample prediction, achieving the highest AUC and KS statistic, indicating stronger discrimination for defaulting loans.

Several questions remain for future work. First, the graph is currently constructed from contemporaneous industrial composition; incorporating forward-looking information, such as announced plant openings and closures, or patent filings, could improve both the timing and the accuracy of graph updates. Second, edge formation and dissolution could be modelled explicitly using dynamic stochastic block models or temporal exponential random graph models, which would provide edge-level uncertainty and distinguish edge disappearance from edge weakening. Third, the portfolio-level consequences of an evolving industrial dependence deserve further study, including tail risk, diversification benefits, concentration risk, and stress testing in which sectoral shocks propagate through the evolving dependence.

\appendix

\section{Mathematical Proofs}
\label{app:proofs}

\begin{lemma}[Positive semidefiniteness and null space of the weighted graph Laplacian]
\label{lem:laplacian_psd}

Fix a year \(t\) and specification \(g\in\{\mathrm S,\mathrm D\}\). Suppose that \(\bm W_t^{(g)}\) is symmetric, has non-negative entries, and has zero diagonal, and let
\begin{equation}
\bm L_t^{(g)}
=
\bm D_t^{(g)}
-
\bm W_t^{(g)}
\end{equation}
be the corresponding weighted graph Laplacian, where
\begin{equation}
\bm D_t^{(g)}
=
\operatorname{diag}
\left(
\bm W_t^{(g)}\bm 1_A
\right).
\end{equation}
Then
\begin{equation}
\bm x^\top
\bm L_t^{(g)}
\bm x
=
\sum_{a<b}
w_{ab,t}^{(g)}
\left(
x_a-x_b
\right)^2,
\end{equation}
and \(\bm L_t^{(g)}\) is symmetric positive semidefinite.
If the weighted graph is connected\footnote{The connectivity conditions are satisfied for every graph used.}, then
\begin{equation}
\operatorname{rank}
\left(
\bm L_t^{(g)}
\right)
=
A-1
\end{equation}
and
\begin{equation}
\operatorname{null}
\left(
\bm L_t^{(g)}
\right)
=
\operatorname{span}\{\bm 1_A\}.
\end{equation}

\end{lemma}

\begin{proof}

For any $\bm x=(x_1,\ldots,x_A)^\top\in\mathbb R^A$,
we have
\begin{align*}
\bm x^\top
\bm L_t^{(g)}
\bm x
&=
\bm x^\top
\bm D_t^{(g)}
\bm x
-
\bm x^\top
\bm W_t^{(g)}
\bm x                                                    \\
&=
\sum_{a=1}^{A}
\left(
\sum_{b=1}^{A}
w_{ab,t}^{(g)}
\right)x_a^2
-
\sum_{a=1}^{A}\sum_{b=1}^{A}
w_{ab,t}^{(g)}
x_a x_b.
\end{align*}
Using
$w_{ab,t}^{(g)}
=
w_{ba,t}^{(g)}$, this expression can be written as
\begin{align*}
\bm x^\top
\bm L_t^{(g)}
\bm x
&=
\frac12
\sum_{a=1}^{A}
\sum_{b=1}^{A}
w_{ab,t}^{(g)}
(x_a-x_b)^2                                             \\
&=
\sum_{a<b}
w_{ab,t}^{(g)}
(x_a-x_b)^2
\geq0.
\end{align*}
Hence
\[
\bm L_t^{(g)}
\succeq0.
\]

Since the diagonal elements of
\(\bm D_t^{(g)}\) are the row sums of
\(\bm W_t^{(g)}\). 
Thus,
\[
\bm L_t^{(g)}\bm 1_A
=
\left(
\bm D_t^{(g)}
-
\bm W_t^{(g)}
\right)\bm 1_A
=
\bm0,
\]
Thus $\bm 1_A
\in
\operatorname{null}
\left(
\bm L_t^{(g)}
\right)$.

Suppose that the graph is connected and that $\bm x^\top
\bm L_t^{(g)}
\bm x
=
0$.
Since every term in
\[
\sum_{a<b}
w_{ab,t}^{(g)}
(x_a-x_b)^2
\]
is non-negative, equality can hold only if $x_a=x_b$
for every pair of areas connected by a positive-weight edge.
Connectivity then implies $x_1=\cdots=x_A$. Therefore $\bm x=c\bm 1_A$ for some scalar \(c\), and hence
\[
\operatorname{null}
\left(
\bm L_t^{(g)}
\right)
=
\operatorname{span}\{\bm 1_A\}.
\]
It follows that the null space has dimension one and therefore
\[
\operatorname{rank}
\left(
\bm L_t^{(g)}
\right)
=
A-1.
\]
\end{proof}

\begin{proposition}[Generalised-variance normalisation of graphs]
\label{prop:graph_scaling}

Fix a year \(t\) and specification
\(g\in\{\mathrm S,\mathrm D\}\), and suppose that the weighted
industry-similarity graph is connected\footnote{The connectivity conditions are satisfied for every graph used.}.

Let
\begin{equation}
c_t^{(g)}
=
\exp
\left\{
\frac{1}{A}
\sum_{a=1}^{A}
\log
\left[
\left\{
\left(
\bm L_t^{(g)}
\right)^{-}
\right\}_{aa}
\right]
\right\},
\label{si_eq:scaled_graph_structure}
\end{equation}
 and define
\begin{equation}
\widetilde{\bm L}_t^{(g)}
=
c_t^{(g)}
\bm L_t^{(g)}.
\end{equation}
Then
\begin{equation}
\left[
\prod_{a=1}^{A}
\left\{
\left(
\widetilde{\bm L}_t^{(g)}
\right)^{-}
\right\}_{aa}
\right]^{1/A}
=
1.
\end{equation}
Thus the generalised marginal variances associated with
\(\widetilde{\bm L}_t^{(g)}\) have unit geometric mean, so the
graphs are placed on a common scale. Furthermore,
\(\widetilde{\bm L}_t^{(g)}\) has the same null space and the same graph
topology as
\(\bm L_t^{(g)}\).
\end{proposition}

\begin{proof}

By Lemma~\ref{lem:laplacian_psd},
\(\bm L_t^{(g)}\) is symmetric positive semidefinite with exactly one zero eigenvalue.

Let its spectral decomposition be
\[
\bm L_t^{(g)}
=
\bm U
\operatorname{diag}
(0,\nu_2,\ldots,\nu_A)
\bm U^\top,
\qquad
\nu_j>0,
\quad
j=2,\ldots,A.
\]
Its Moore-Penrose generalised inverse is therefore
\[
\left(
\bm L_t^{(g)}
\right)^{-}
=
\bm U
\operatorname{diag}
(0,\nu_2^{-1},\ldots,\nu_A^{-1})
\bm U^\top.
\]
For every \(a\),
\[
\left\{
\left(
\bm L_t^{(g)}
\right)^{-}
\right\}_{aa}
=
\sum_{j=2}^{A}
\nu_j^{-1}U_{aj}^{2}
>0.
\]
Hence the logarithms in Eq.\eqref{si_eq:scaled_graph_structure} are well defined and $c_t^{(g)}>0$.

Since multiplication of a matrix by a nonzero scalar rescales its Moore-Penrose generalised inverse by the reciprocal scalar,
\[
\left(
c_t^{(g)}
\bm L_t^{(g)}
\right)^{-}
=
\frac{1}{c_t^{(g)}}
\left(
\bm L_t^{(g)}
\right)^{-}.
\]
Since
\[
\widetilde{\bm L}_t^{(g)}
=
c_t^{(g)}
\bm L_t^{(g)},
\]
it follows that
\[
\left(
\widetilde{\bm L}_t^{(g)}
\right)^{-}
=
\frac{1}{c_t^{(g)}}
\left(
\bm L_t^{(g)}
\right)^{-}.
\]
Therefore,
\begin{align*}
&
\left[
\prod_{a=1}^{A}
\left\{
\left(
\widetilde{\bm L}_t^{(g)}
\right)^{-}
\right\}_{aa}
\right]^{1/A}
\\
&\qquad=
\left[
\prod_{a=1}^{A}
\frac{1}{c_t^{(g)}}
\left\{
\left(
\bm L_t^{(g)}
\right)^{-}
\right\}_{aa}
\right]^{1/A}
\\
&\qquad=
\frac{1}{c_t^{(g)}}
\left[
\prod_{a=1}^{A}
\left\{
\left(
\bm L_t^{(g)}
\right)^{-}
\right\}_{aa}
\right]^{1/A}.
\end{align*}

On the other hand, the definition in
Eq.\eqref{si_eq:scaled_graph_structure} gives
\begin{align*}
c_t^{(g)}
&=
\exp
\left\{
\frac{1}{A}
\sum_{a=1}^{A}
\log
\left[
\left\{
\left(
\bm L_t^{(g)}
\right)^{-}
\right\}_{aa}
\right]
\right\}
\\
&=
\left[
\prod_{a=1}^{A}
\left\{
\left(
\bm L_t^{(g)}
\right)^{-}
\right\}_{aa}
\right]^{1/A}.
\end{align*}
Substituting this expression into the preceding equality yields
\[
\left[
\prod_{a=1}^{A}
\left\{
\left(
\widetilde{\bm L}_t^{(g)}
\right)^{-}
\right\}_{aa}
\right]^{1/A}
=
\frac{c_t^{(g)}}
{c_t^{(g)}}
=
1.
\]

Thus, after scaling, the geometric mean of the generalised marginal variances is exactly one.  Since the same normalisation is applied to every year and to both specifications, differences in the scale of the original graph Laplacians are removed before the matrices enter the model. Therefore, variation across \(\widetilde{\bm L}_t^{(g)}\) reflects differences in the standardised graphs rather than arbitrary differences in the overall generalised-variance scale.

Besides, since $c_t^{(g)}>0$,
we have
\[
\widetilde{\bm L}_t^{(g)}\bm x=\bm0
\quad\Longleftrightarrow\quad
\bm L_t^{(g)}\bm x=\bm0.
\]
Hence
\[
\operatorname{null}
\left(
\widetilde{\bm L}_t^{(g)}
\right)
=
\operatorname{null}
\left(
\bm L_t^{(g)}
\right).
\]
Moreover, multiplication by a strictly positive scalar leaves the locations of all zero and nonzero matrix entries unchanged. Therefore, the scaling preserves the graph topology as well as the
null space.
\end{proof}

\begin{proposition}[Positive Definiteness of Precision Matrices and Proper Gaussian Priors]
\label{prop:innovation_properness}

For every year \(t\), specification
\(g\in\{\mathrm S,\mathrm D\}\), and
\(\lambda\in[0,1)\), define
\begin{equation}
\widehat{\bm L}_t^{(g)}(\lambda)
=
(1-\lambda)I_A
+
\lambda
\widetilde{\bm L}_t^{(g)},
\end{equation} 
and
\begin{equation}
\bm R_t^{(g)}(\lambda)
=
r_t^{(g)}(\lambda)
\widehat{\bm L}_t^{(g)}(\lambda),
\label{si_eq:annual_graph_precision}
\end{equation} 
where
\(r_t^{(g)}(\lambda)\) is defined as
\begin{equation}
r_t^{(g)}(\lambda)
=
\exp
\left\{
\frac{1}{A}
\sum_{a=1}^{A}
\log
\left[
\left\{
\widehat{\bm L}_t^{(g)}(\lambda)^{-1}
\right\}_{aa}
\right]
\right\}.
\label{si_eq:r_definition}
\end{equation}

Then both
\(\widehat{\bm L}_t^{(g)}(\lambda)\) and
\(\bm R_t^{(g)}(\lambda)\) are symmetric positive definite.
Consequently, the static latent-effect distribution
\begin{equation}
\bm u
\mid
\tau_u,\lambda
\sim
\mathcal N_A
\left[
\bm0,
\left\{
\tau_u
\bm R_{t_0}^{(\mathrm S)}(\lambda)
\right\}^{-1}
\right],
\end{equation}
and the dynamic innovation distribution
\begin{equation}
\bm\varepsilon_t
\mid
\tau_\phi,\lambda
\sim
\mathcal N_A
\left[
\bm0,
\left\{
\tau_\phi
\bm R_t^{(\mathrm D)}(\lambda)
\right\}^{-1}
\right]
\end{equation}
define proper multivariate Gaussian distributions for
\(\tau_u>0\) and \(\tau_\phi>0\), respectively.
\end{proposition}

\begin{proof}

Fix \(t\), \(g\), and \(\lambda\in[0,1)\). By
Lemma~\ref{lem:laplacian_psd} and Proposition~\ref{prop:graph_scaling},
\(\widetilde{\bm L}_t^{(g)}\) is symmetric positive semidefinite.
Hence, for every
\(\bm x\in\mathbb R^A\),
\[
\bm x^\top
\widetilde{\bm L}_t^{(g)}
\bm x
\geq0.
\]

Now let \(\bm x\neq\bm0\). Then
\begin{align*}
\bm x^\top
\widehat{\bm L}_t^{(g)}(\lambda)
\bm x
&=
(1-\lambda)
\bm x^\top\bm x
+
\lambda
\bm x^\top
\widetilde{\bm L}_t^{(g)}
\bm x                                                     \\
&\geq
(1-\lambda)\|\bm x\|^2.
\end{align*}
Since $0\leq\lambda<1$, we have $1-\lambda>0$.

Therefore, for every nonzero \(\bm x\),
\[
\bm x^\top
\widehat{\bm L}_t^{(g)}(\lambda)
\bm x
>
0.
\]
Thus
\[
\widehat{\bm L}_t^{(g)}(\lambda)
\succ0.
\]

Since \(\widehat{\bm L}_t^{(g)}(\lambda)\) is symmetric positive definite, its inverse is also symmetric positive definite. In particular,
\[
\left\{
\widehat{\bm L}_t^{(g)}(\lambda)^{-1}
\right\}_{aa}
>0,
\qquad
a=1,\ldots,A.
\]
Hence every logarithm in Eq.\eqref{si_eq:r_definition}
is well defined, and
\[
0
<
r_t^{(g)}(\lambda)
<
\infty.
\]
Since multiplication by a strictly positive scalar preserves positive
definiteness,
\[
\bm R_t^{(g)}(\lambda)
=
r_t^{(g)}(\lambda)
\widehat{\bm L}_t^{(g)}(\lambda)
\succ0.
\]

It follows in particular that
\[
\tau_u
\bm R_{t_0}^{(\mathrm S)}(\lambda)
\succ0
\qquad\text{and}\qquad
\tau_\phi
\bm R_t^{(\mathrm D)}(\lambda)
\succ0
\]
for every \(\tau_u>0\) and \(\tau_\phi>0\). Their inverses therefore
exist and are positive definite covariance matrices, which proves that
the static latent-effect distribution and the dynamic innovation
distribution are proper multivariate Gaussian distributions.
\end{proof}

\begin{proposition}[Scale interpretation of the precision parameters]
\label{prop:tau_interpretation}

For every year \(t\), specification
\(g\in\{\mathrm S,\mathrm D\}\), and
\(\lambda\in[0,1)\), then
\begin{equation}
\left[
\prod_{a=1}^{A}
\left\{
\left(
\bm R_t^{(g)}(\lambda)
\right)^{-1}
\right\}_{aa}
\right]^{1/A}
=
1.
\end{equation}

Consequently, under the static latent-effect distribution,
\begin{equation}
\bm u
\mid
\tau_u,\lambda
\sim
\mathcal N_A
\left[
\bm0,
\left\{
\tau_u
\bm R_{t_0}^{(\mathrm S)}(\lambda)
\right\}^{-1}
\right],
\end{equation}
the marginal variances satisfy
\begin{equation}
\left[
\prod_{a=1}^{A}
\operatorname{Var}
\left(
u_a
\mid
\tau_u,\lambda
\right)
\right]^{1/A}
=
\tau_u^{-1}.
\end{equation}
Likewise, under the dynamic annual innovation distribution,
\begin{equation}
\bm\varepsilon_t
\mid
\tau_\phi,\lambda
\sim
\mathcal N_A
\left[
\bm0,
\left\{
\tau_\phi
\bm R_t^{(\mathrm D)}(\lambda)
\right\}^{-1}
\right],
\end{equation}
the marginal innovation variances satisfy
\begin{equation}
\left[
\prod_{a=1}^{A}
\operatorname{Var}
\left(
\varepsilon_{a,t}
\mid
\tau_\phi,\lambda
\right)
\right]^{1/A}
=
\tau_\phi^{-1}.
\end{equation}
Consequently,
\(\tau_u^{-1/2}\) and \(\tau_\phi^{-1/2}\) have the corresponding
geometric-mean marginal standard-deviation interpretations for the
static latent effect and the dynamic annual innovations, respectively.
\end{proposition}

\begin{proof}

Fix \(t\), \(g\), and \(\lambda\). By Eq.\eqref{si_eq:annual_graph_precision},
\[
\bm R_t^{(g)}(\lambda)
=
r_t^{(g)}(\lambda)
\widehat{\bm L}_t^{(g)}(\lambda).
\]
Since
\(r_t^{(g)}(\lambda)>0\),
\[
\left\{
\bm R_t^{(g)}(\lambda)
\right\}^{-1}
=
\frac{1}
{r_t^{(g)}(\lambda)}
\left\{
\widehat{\bm L}_t^{(g)}(\lambda)
\right\}^{-1}.
\]
Therefore,
\begin{align*}
&
\left[
\prod_{a=1}^{A}
\left\{
\left(
\bm R_t^{(g)}(\lambda)
\right)^{-1}
\right\}_{aa}
\right]^{1/A}
\\
&\qquad=
\left[
\prod_{a=1}^{A}
\frac{1}
{r_t^{(g)}(\lambda)}
\left\{
\left(
\widehat{\bm L}_t^{(g)}(\lambda)
\right)^{-1}
\right\}_{aa}
\right]^{1/A}
\\
&\qquad=
\frac{1}
{r_t^{(g)}(\lambda)}
\left[
\prod_{a=1}^{A}
\left\{
\left(
\widehat{\bm L}_t^{(g)}(\lambda)
\right)^{-1}
\right\}_{aa}
\right]^{1/A}.
\end{align*}

From the definition of \(r_t^{(g)}(\lambda)\) in
Eq.\eqref{si_eq:r_definition},
\begin{align*}
r_t^{(g)}(\lambda)
&=
\exp
\left\{
\frac{1}{A}
\sum_{a=1}^{A}
\log
\left[
\left\{
\widehat{\bm L}_t^{(g)}(\lambda)^{-1}
\right\}_{aa}
\right]
\right\}
\\
&=
\left[
\prod_{a=1}^{A}
\left\{
\left(
\widehat{\bm L}_t^{(g)}(\lambda)
\right)^{-1}
\right\}_{aa}
\right]^{1/A}.
\end{align*}
Substituting this expression into the preceding result gives
\[
\left[
\prod_{a=1}^{A}
\left\{
\left(
\bm R_t^{(g)}(\lambda)
\right)^{-1}
\right\}_{aa}
\right]^{1/A}
=
1.
\]

For the static specification,
\[
\operatorname{Cov}
\left(
\bm u
\mid
\tau_u,\lambda
\right)
=
\tau_u^{-1}
\left\{
\bm R_{t_0}^{(\mathrm S)}(\lambda)
\right\}^{-1}.
\]
Hence
\[
\left[
\prod_{a=1}^{A}
\operatorname{Var}
\left(
u_a
\mid
\tau_u,\lambda
\right)
\right]^{1/A}
=
\tau_u^{-1}.
\]

For the dynamic specification,
\[
\operatorname{Cov}
\left(
\bm\varepsilon_t
\mid
\tau_\phi,\lambda
\right)
=
\tau_\phi^{-1}
\left\{
\bm R_t^{(\mathrm D)}(\lambda)
\right\}^{-1},
\]
and therefore
\[
\left[
\prod_{a=1}^{A}
\operatorname{Var}
\left(
\varepsilon_{a,t}
\mid
\tau_\phi,\lambda
\right)
\right]^{1/A}
=
\tau_\phi^{-1}.
\]

Taking square roots of the strictly positive marginal variances gives
the corresponding geometric-mean marginal standard deviations
\(\tau_u^{-1/2}\) and \(\tau_\phi^{-1/2}\).
\end{proof}

\begin{proposition}[Unified distribution of the transformed innovations]
\label{prop:unified_innovation}

Suppose that the initial dynamic effect satisfies
\begin{equation}
\bm\phi_1
\mid
\tau_\phi,\rho,\lambda
\sim
\mathcal N_A
\left[
\bm0,
\left\{
\tau_\phi(1-\rho^2)
\bm R_1^{(\mathrm D)}(\lambda)
\right\}^{-1}
\right],
\end{equation}
and, for \(t=2,\ldots,T\), the state process follows
\begin{equation}
\bm\phi_t
=
\rho\bm\phi_{t-1}
+
\bm\varepsilon_t,
\qquad
t=2,\ldots,T,
\qquad
|\rho|<1.
\label{si_eq:industry_ar1_state}
\end{equation}
with innovations distributed as 
\begin{equation}
\bm\varepsilon_t
\mid
\tau_\phi,\lambda
\sim
\mathcal N_A
\left[
\bm0,
\left\{
\tau_\phi
\bm R_t^{(\mathrm D)}(\lambda)
\right\}^{-1}
\right],
\qquad
t=2,\ldots,T.
\label{si_eq:annual_innovation_distribution}
\end{equation}
Assume that, conditional on the model parameters, the initial latent effect and all subsequent innovations are mutually independent.

Define
\[
\bm\varepsilon_1
=
\sqrt{1-\rho^2}\,\bm\phi_1,
\qquad
\bm\varepsilon_t
=
\bm\phi_t-\rho\bm\phi_{t-1},
\quad
t=2,\ldots,T.
\]
Then
\[
\bm\varepsilon_t
\mid
\tau_\phi,\lambda
\sim
\mathcal N_A
\left[
\bm0,
\left\{
\tau_\phi
\bm R_t^{(\mathrm D)}(\lambda)
\right\}^{-1}
\right],
\qquad
t=1,\ldots,T.
\]
Thus the initial transformed innovation and all subsequent innovations are expressed on exactly the unified distribution. Moreover, conditional on the parameters, the transformed innovations are mutually independent across years.
\end{proposition}

\begin{proof}

For \(t=2,\ldots,T\), $\bm\varepsilon_t=\bm\phi_t-\rho\bm\phi_{t-1}$ is precisely the innovation in Eq.\eqref{si_eq:industry_ar1_state}. Its distribution is therefore, by Eq.\eqref{si_eq:annual_innovation_distribution},
\[
\bm\varepsilon_t
\mid
\tau_\phi,\lambda
\sim
\mathcal N_A
\left[
\bm0,
\left\{
\tau_\phi
\bm R_t^{(\mathrm D)}(\lambda)
\right\}^{-1}
\right].
\]

It remains to establish the result for \(t=1\).
From Eq.\eqref{eq:industry_initial_distribution},
\begin{align*}
\operatorname{Cov}
\left(
\bm\phi_1
\mid
\tau_\phi,\rho,\lambda
\right)
&=
\left\{
\tau_\phi(1-\rho^2)
\bm R_1^{(\mathrm D)}(\lambda)
\right\}^{-1}                                                     \\
&=
(1-\rho^2)^{-1}
\left\{
\tau_\phi
\bm R_1^{(\mathrm D)}(\lambda)
\right\}^{-1}.
\end{align*}
A deterministic linear transformation of a Gaussian vector remains
Gaussian. Therefore,
$\bm\varepsilon_1
=
\sqrt{1-\rho^2}\,\bm\phi_1$ is Gaussian with mean zero and covariance
\begin{align*}
\operatorname{Cov}
\left(
\bm\varepsilon_1
\mid
\tau_\phi,\rho,\lambda
\right)
&=
(1-\rho^2)
\operatorname{Cov}
\left(
\bm\phi_1
\mid
\tau_\phi,\rho,\lambda
\right)                                                          \\
&=
(1-\rho^2)
(1-\rho^2)^{-1}
\left\{
\tau_\phi
\bm R_1^{(\mathrm D)}(\lambda)
\right\}^{-1}                                                     \\
&=
\left\{
\tau_\phi
\bm R_1^{(\mathrm D)}(\lambda)
\right\}^{-1}.
\end{align*}
Hence
\[
\bm\varepsilon_1
\mid
\tau_\phi,\lambda
\sim
\mathcal N_A
\left[
\bm0,
\left\{
\tau_\phi
\bm R_1^{(\mathrm D)}(\lambda)
\right\}^{-1}
\right].
\]
The dependence on \(\rho\) disappears after the transformation. 

Finally, by the mutual conditional independence
assumed in the model construction,
\[
\bm\phi_1
\indep
(\bm\varepsilon_2,\ldots,\bm\varepsilon_T)
\mid
\tau_\phi,\rho,\lambda.
\]
Since
$\bm\varepsilon_1
=\sqrt{1-\rho^2}\bm\phi_1$
is a deterministic transformation of
$\bm\phi_1$ conditional on the model parameters,
it remains independent of the subsequent
innovations. Together with their assumed
mutual conditional independence, this establishes
the mutual conditional independence of
$\bm\varepsilon_1,\ldots,\bm\varepsilon_T$.

\end{proof}

\begin{lemma}[Stacked innovation transformation]
\label{lem:stacked_innovation_transformation}

Let $\bm\phi=\left(\bm\phi_1^\top,\ldots,\bm\phi_T^\top\right)^\top\in\mathbb R^{AT}$, and $\bm\varepsilon
=\left(\bm\varepsilon_1^\top,\ldots,\bm\varepsilon_T^\top\right)^\top\in\mathbb R^{AT}$, where $\bm\varepsilon_1=\sqrt{1-\rho^2}\,\bm\phi_1$ and $\bm\varepsilon_t=\bm\phi_t-\rho\bm\phi_{t-1}, t=2,\ldots,T$.

Define
\begin{equation}
B_\rho
=
\begin{pmatrix}
\sqrt{1-\rho^2} & 0 & \cdots & 0 \\
-\rho & 1 & \ddots & \vdots \\
0 & \ddots & \ddots & 0 \\
0 & \cdots & -\rho & 1
\end{pmatrix},
\label{si_eq:temporal_innovation_matrix}
\end{equation}
and
\begin{equation}
\mathbb B_\rho
=
B_\rho\otimes I_A.
\end{equation}
Then
\begin{equation}
\bm\varepsilon
=
\mathbb B_\rho\bm\phi.
\label{si_eq:transformed_innovation}
\end{equation}
\end{lemma}

\begin{proof}

By the definition of the Kronecker product,
\[
\mathbb B_\rho
=
B_\rho\otimes I_A
=
\begin{pmatrix}
\sqrt{1-\rho^2}I_A
&
0
&
\cdots
&
0
\\
-\rho I_A
&
I_A
&
\ddots
&
\vdots
\\
0
&
\ddots
&
\ddots
&
0
\\
0
&
\cdots
&
-\rho I_A
&
I_A
\end{pmatrix}.
\]
Therefore,
\begin{align*}
\mathbb B_\rho\bm\phi
&=
\begin{pmatrix}
\sqrt{1-\rho^2}I_A
&
0
&
\cdots
&
0
\\
-\rho I_A
&
I_A
&
\ddots
&
\vdots
\\
0
&
\ddots
&
\ddots
&
0
\\
0
&
\cdots
&
-\rho I_A
&
I_A
\end{pmatrix}
\begin{pmatrix}
\bm\phi_1\\
\bm\phi_2\\
\vdots\\
\bm\phi_T
\end{pmatrix}
=
\begin{pmatrix}
\sqrt{1-\rho^2}\,\bm\phi_1
\\
\bm\phi_2-\rho\bm\phi_1
\\
\bm\phi_3-\rho\bm\phi_2
\\
\vdots
\\
\bm\phi_T-\rho\bm\phi_{T-1}
\end{pmatrix}.
\end{align*}
By the definition of the transformed innovations, $\bm\varepsilon_1=\sqrt{1-\rho^2}\,\bm\phi_1$ and $\bm\varepsilon_t=\bm\phi_t-\rho\bm\phi_{t-1},t=2,\ldots,T$.

Hence
\[
\mathbb B_\rho\bm\phi
=
\begin{pmatrix}
\bm\varepsilon_1\\
\bm\varepsilon_2\\
\vdots\\
\bm\varepsilon_T
\end{pmatrix}
=
\bm\varepsilon.
\]
Therefore,
\[\bm\varepsilon
=
\mathbb B_\rho\bm\phi.
\]
\end{proof}

\begin{proposition}[Joint precision matrix representation]
\label{prop:joint_precision}

Let $\bm\phi=(\bm\phi_1^\top,\ldots,\bm\phi_T^\top)^\top\in\mathbb R^{AT}$, and define $\mathbb B_\rho=B_\rho\otimes I_A$, where \(B_\rho\) is given in Eq.\eqref{si_eq:transformed_innovation}. Let
\begin{equation}
\mathcal R^{(\mathrm D)}(\lambda)
=
\operatorname{blockdiag}
\left\{
\bm R_1^{(\mathrm D)}(\lambda),
\ldots,
\bm R_T^{(\mathrm D)}(\lambda)
\right\}.
\end{equation}
Then the unconstrained dynamic latent industrial effect is jointly Gaussian with
precision matrix
\begin{equation}
\bm Q_\phi^{(\mathrm D)}
=
\tau_\phi
\mathbb B_\rho^\top
\mathcal R^{(\mathrm D)}(\lambda)
\mathbb B_\rho.
\label{si_eq:joint_industry_precision}
\end{equation}
\end{proposition}

\begin{proof}

By Proposition~\ref{prop:unified_innovation}, the stacked transformed innovation vector $\bm\varepsilon=(\bm\varepsilon_1^\top,\ldots,\bm\varepsilon_T^\top)^\top$ has conditional distribution
\[
\bm\varepsilon
\mid
\tau_\phi,\lambda
\sim
\mathcal N_{AT}
\left[
\bm0,
\left\{
\tau_\phi
\mathcal R^{(\mathrm D)}(\lambda)
\right\}^{-1}
\right],
\]
and by Lemma~\ref{lem:stacked_innovation_transformation},
\[
\bm\varepsilon
=
\mathbb B_\rho\bm\phi.
\]
Since \(|\rho|<1\), \(B_\rho\) is lower triangular with diagonal
entries
\[
\sqrt{1-\rho^2},1,\ldots,1,
\]
all of which are strictly positive.  Hence \(B_\rho\), and therefore
\(\mathbb B_\rho=B_\rho\otimes I_A\), is nonsingular.  It follows that
\[
\bm\phi
=
\mathbb B_\rho^{-1}\bm\varepsilon
\]
is Gaussian.  Its covariance matrix is
\begin{align*}
\operatorname{Cov}(\bm\phi)
&=
\mathbb B_\rho^{-1}
\left\{
\tau_\phi
\mathcal R^{(\mathrm D)}(\lambda)
\right\}^{-1}
\mathbb B_\rho^{-\top}.
\end{align*}
Taking the inverse gives
\begin{align*}
\operatorname{Prec}(\bm\phi)=\bm Q_\phi^{(\mathrm D)}
&=
\mathbb B_\rho^\top
\left\{
\tau_\phi
\mathcal R^{(\mathrm D)}(\lambda)
\right\}
\mathbb B_\rho                                                     \\
&=
\tau_\phi
\mathbb B_\rho^\top
\mathcal R^{(\mathrm D)}(\lambda)
\mathbb B_\rho,
\end{align*}
which proves Eq.\eqref{si_eq:joint_industry_precision}.
\end{proof}

\begin{proposition}[Dynamic nonseparability]
\label{prop:dynamic_nonseparability}

Under the dynamic specification, with the normalisation in Section~\ref{sec:graph_scaling} in the main text, the joint precision matrix is separable in the form
\[
\bm Q_\phi^{(\mathrm D)}
=
\bm K\otimes\bm S,
\]
for positive-definite matrices
\(\bm K\in\mathbb R^{T\times T}\) and
\(\bm S\in\mathbb R^{A\times A}\),
if and only if
\[
\bm R_1^{(\mathrm D)}(\lambda)
=
\cdots
=
\bm R_T^{(\mathrm D)}(\lambda).
\]
Consequently, if there exist two years \(s\) and \(t\) such
that
\[
\bm R_s^{(\mathrm D)}(\lambda)
\neq
\bm R_t^{(\mathrm D)}(\lambda),
\]
then
\(\bm Q_\phi^{(\mathrm D)}\)
is nonseparable.
\end{proposition}

\begin{proof}

Write $\bm R_t=\bm R_t^{(\mathrm D)}(\lambda),t=1,\ldots,T$.
By Proposition~\ref{prop:joint_precision},
\[
\bm Q_\phi^{(\mathrm D)}
=
\tau_\phi
\mathbb B_\rho^\top
\operatorname{blockdiag}
\left(
\bm R_1,\ldots,\bm R_T
\right)
\mathbb B_\rho,
\]
where $\mathbb B_\rho=B_\rho\otimes I_A$.
Direct block multiplication gives
\[
\tau_\phi^{-1}\bm Q_\phi^{(\mathrm D)}
=
\begin{pmatrix}
(1-\rho^2)\bm R_1+\rho^2\bm R_2
&
-\rho\bm R_2
&
0
&
\cdots
&
0
\\
-\rho\bm R_2
&
\bm R_2+\rho^2\bm R_3
&
-\rho\bm R_3
&
\ddots
&
\vdots
\\
0
&
-\rho\bm R_3
&
\bm R_3+\rho^2\bm R_4
&
\ddots
&
0
\\
\vdots
&
\ddots
&
\ddots
&
\ddots
&
-\rho\bm R_T
\\
0
&
\cdots
&
0
&
-\rho\bm R_T
&
\bm R_T
\end{pmatrix}.
\label{eq:dynamic_precision_block_form}
\]

Suppose that
\[
\bm Q_\phi^{(\mathrm D)}
=
\bm K\otimes\bm S
\]
for positive-definite matrices
\(\bm K\in\mathbb R^{T\times T}\) and
\(\bm S\in\mathbb R^{A\times A}\).
By the definition of the Kronecker product, the
\((i,j)\)th \(A\times A\) block of
\(\bm K\otimes\bm S\) is
\[
K_{ij}\bm S.
\]

First suppose that $\rho\neq0$. For \(t=2,\ldots,T\), comparison of the \((t,t-1)\) blocks of $\tau_\phi^{-1}\bm Q_\phi^{(\mathrm D)}$ and \(\bm K\otimes\bm S\) gives
\[
-\tau_\phi\rho\bm R_t
=
K_{t,t-1}\bm S.
\]
Since
\(\tau_\phi>0\)
and
\(\rho\neq0\),
\[
\bm R_t
=
-\frac{K_{t,t-1}}
{\tau_\phi\rho}
\bm S,
\qquad
t=2,\ldots,T.
\]
Since both
\(\bm R_t\)
and
\(\bm S\)
are positive definite, $-\frac{K_{t,t-1}}{\tau_\phi\rho}>0$.

Hence
\[
\bm R_t
\propto
\bm S,
\qquad
t=2,\ldots,T.
\]

For the first diagonal block,
\[
\tau_\phi
\left[
(1-\rho^2)\bm R_1
+
\rho^2\bm R_2
\right]
=
K_{11}\bm S.
\]
Since
\(\bm R_2\propto\bm S\),
there exists
\(a_2>0\)
such that $\bm R_2=a_2\bm S$.

Therefore, $\tau_\phi(1-\rho^2)\bm R_1=\left(K_{11}-\tau_\phi\rho^2a_2\right)\bm S$.
Since $|\rho|<1$, we have $1-\rho^2>0$,

and therefore
\[
\bm R_1
=
\frac{
K_{11}-\tau_\phi\rho^2a_2
}{
\tau_\phi(1-\rho^2)
}
\bm S.
\]
Hence
\[
\bm R_1
\propto
\bm S.
\]
Thus
\[
\bm R_t
\propto
\bm S,
\qquad
t=1,\ldots,T.
\]

Now suppose that $\rho=0$. Then $\tau_\phi^{-1}\bm Q_\phi^{(\mathrm D)}$ reduces to
\[
\bm Q_\phi^{(\mathrm D)}
=
\tau_\phi
\operatorname{blockdiag}
\left(
\bm R_1,\ldots,\bm R_T
\right).
\]
Since
\[
\bm Q_\phi^{(\mathrm D)}
=
\bm K\otimes\bm S,
\]
comparison of the diagonal blocks gives
\[
\tau_\phi\bm R_t
=
K_{tt}\bm S,
\qquad
t=1,\ldots,T.
\]
Therefore,
\[
\bm R_t
=
\frac{K_{tt}}{\tau_\phi}\bm S,
\qquad
t=1,\ldots,T.
\]
Hence, also when
\(\rho=0\),
\[
\bm R_t
\propto
\bm S,
\qquad
t=1,\ldots,T.
\]

Therefore, for every \(|\rho|<1\), separability implies that all annual matrices are pairwise proportional. Thus, for any
\(s,t\), there exists
\(c_{ts}>0\)
such that
\[
\bm R_t
=
c_{ts}\bm R_s,
\qquad
\bm R_t^{-1}
=
c_{ts}^{-1}\bm R_s^{-1}
\]

By Section~\ref{sec:graph_scaling} in the main text,
\[
\bm R_t
=
r_t^{(\mathrm D)}(\lambda)
\widehat{\bm L}_t^{(\mathrm D)}(\lambda),
\]
with
\[
r_t^{(\mathrm D)}(\lambda)
=
\left[
\prod_{a=1}^{A}
\left\{
\left(
\widehat{\bm L}_t^{(\mathrm D)}(\lambda)
\right)^{-1}
\right\}_{aa}
\right]^{1/A}.
\]
Hence
\begin{align*}
&
\left[
\prod_{a=1}^{A}
\left\{
\left(
\bm R_t
\right)^{-1}
\right\}_{aa}
\right]^{1/A}
\\
&=
\frac{1}
{r_t^{(\mathrm D)}(\lambda)}
\left[
\prod_{a=1}^{A}
\left\{
\left(
\widehat{\bm L}_t^{(\mathrm D)}(\lambda)
\right)^{-1}
\right\}_{aa}
\right]^{1/A}
\\
&=
1.
\end{align*}
Therefore,
\[
\left[
\prod_{a=1}^{A}
\left\{
(\bm R_t^{-1})_{aa}
\right\}
\right]^{1/A}
=
1.
\]

Thus,
\begin{align*}
1
&=
\left[
\prod_{a=1}^{A}
\left\{
(\bm R_t^{-1})_{aa}
\right\}
\right]^{1/A}
\\
&=
\left[
\prod_{a=1}^{A}
c_{ts}^{-1}
\left\{
(\bm R_s^{-1})_{aa}
\right\}
\right]^{1/A}
\\
&=
c_{ts}^{-1}
\left[
\prod_{a=1}^{A}
\left\{
(\bm R_s^{-1})_{aa}
\right\}
\right]^{1/A}
\\
&=
c_{ts}^{-1}.
\end{align*}
Therefore,
$c_{ts}=1$.
Hence $\bm R_t=\bm R_s$ for every \(s,t\), and consequently
\[
\bm R_1
=
\bm R_2
=
\cdots
=
\bm R_T.
\]
Thus,
\[
\bm Q_\phi^{(\mathrm D)}
\text{ separable}
\quad\Longrightarrow\quad
\bm R_1
=
\cdots
=
\bm R_T.
\]

\vspace{1cm}

\textbf{Conversely}, suppose that $\bm R_1=\cdots=\bm R_T=\bm R$.

Then
\[
\operatorname{blockdiag}
\left(
\bm R_1,\ldots,\bm R_T
\right)
=
I_T\otimes\bm R.
\]
Therefore,
\begin{align*}
\bm Q_\phi^{(\mathrm D)}
&=
\tau_\phi
(B_\rho^\top\otimes I_A)
(I_T\otimes\bm R)
(B_\rho\otimes I_A)
\\
&=
\tau_\phi
(B_\rho^\top I_T B_\rho)
\otimes
(I_A\bm R I_A)
\\
&=
\tau_\phi
(B_\rho^\top B_\rho)
\otimes
\bm R.
\end{align*}
Hence
\[
\bm R_1
=
\cdots
=
\bm R_T
\quad\Longrightarrow\quad
\bm Q_\phi^{(\mathrm D)}
\text{ separable}.
\]

\vspace{1cm}

Combining the two directions,
\[
\bm Q_\phi^{(\mathrm D)}
\text{ is separable}
\quad\Longleftrightarrow\quad
\bm R_1
=
\cdots
=
\bm R_T.
\]
Therefore,
\[
\exists\,s,t
\quad\text{such that}\quad
\bm R_s\neq\bm R_t
\]
implies
\[
\bm Q_\phi^{(\mathrm D)}
\text{ is nonseparable}.
\]
\end{proof}

\section{Prior Specification}
\label{si_sec:prior}

The regression coefficients are assigned the default fixed-effect priors in R-INLA. Specifically, an improper flat prior is adopted for the intercept, \(p(\beta_0)\propto1\), while the remaining regression coefficients are assigned independent zero-mean Gaussian priors,
\[
\beta_j\sim\mathcal N(0,0.001^{-1}),
\qquad j=1,\ldots,p.
\]

The smooth duration effect is represented through a second-order random walk with precision scale parameter \(\tau_f\). We use the penalised complexity prior
\begin{equation}
\Pr(\sigma_f>1)=0.01,
\qquad
\sigma_f=\tau_f^{-1/2}.
\label{eq:prior_duration_pc}
\end{equation}

The structure weight $\lambda$ is represented as
\begin{equation}
\theta_\lambda
\sim
\mathcal N(0,1.5^2),
\qquad
\lambda
=
\operatorname{logit}^{-1}(\theta_\lambda).
\label{eq:prior_lambda_internal}
\end{equation}
The induced prior on \(\lambda\) has median \(0.5\) and an approximate \(95\%\) interval \((0.05,0.95)\), placing relatively little prior probability near the upper boundary. The posterior concentration of \(\lambda\) close to 1 in Table~\ref{tab:parameter_estimates} in the main text, therefore, reflects information provided by data rather than the prior.

The precision scale parameters of the latent effects use the same
Gamma prior family. Specifically, the static industrial effect in \(M_2\), and the dynamic industrial effect in \(M_3\) use
\begin{equation}
\tau_u
\sim
\operatorname{Gamma}(1,0.001),
\qquad
\tau_\phi
\sim
\operatorname{Gamma}(1,0.001),
\label{eq:prior_latent_precision}
\end{equation}

For the dynamic specification, the autoregressive coefficient is represented as
\begin{equation}
\theta_\rho\sim\mathcal N(0,1),
\qquad
\rho=\tanh(\theta_\rho),
\label{eq:prior_rho_internal}
\end{equation}
which guarantees \(|\rho|<1\) while allowing substantial prior support for both positive and negative persistence.

\bibliographystyle{apalike}  
\bibliography{references} 

\end{document}